\documentclass[a4paper,fleqn]{cas-sc}

\usepackage[numbers]{natbib}
\usepackage{url}
\usepackage{hyperref}
\usepackage{multirow}
\usepackage{hyperref}
\usepackage[normalem]{ulem}

\def\tsc#1{\csdef{#1}{\textsc{\lowercase{#1}}\xspace}}
\tsc{WGM}
\tsc{QE}

\newtheorem{theorem}{Theorem}
\newtheorem{corollary}{Corollary}
\newproof{pf}{Proof}
\newproof{pot1}{Proof of Theorem \ref{thm1}}
\newproof{pot2}{Proof of Theorem \ref{thm2}}

\newtheorem{Exp}{Example}
\newtheorem{remark}{Remark}

\renewcommand{\i}{{\mathrm{i}}}

\begin{document}
\newcommand{\alex}[1]{\textcolor{blue}{#1}}
\newcommand{\susanna}[1]{\textcolor{magenta}{#1}}
\newcommand{\anna}[1]{\textcolor{red}{#1}}
\let\WriteBookmarks\relax
\def\floatpagepagefraction{1}
\def\textpagefraction{.001}

\shorttitle{Spectral clustering of stochastic matrices with complex eigenvalues}    

\shortauthors{Frank~\textit{et al}.}  

\title [mode = title]{Spectral clustering of Markov chain transition matrices with complex eigenvalues}

%
\cortext[cor1]{Corresponding author}
\author[1]{Anna-Simone Frank}[type=author,orcid=0000-0002-3728-3476]
\ead{anna-simone.frank@uib.no}
\cormark[1]

\credit{Conceptualization, Methodology, Software, Validation, Formal analysis, Investigation, Writing - original draft, Writing - review \& editing, Visualization}

\author[2]{Alexander Sikorski}[type=author,
       orcid=0000-0001-9051-650X] 
\ead{sikorski@zib.de}
\credit{Methodology, Validation, Formal analysis, Investigation, Writing - review \& editing}


\author[1]{Susanna R\"oblitz}[type=author,
       orcid=0000-0002-2735-0030] 
\ead{susanna.roblitz@uib.no}
\credit{Methodology, Software, Validation, Formal analysis,  Investigation, Writing - original draft, Writing - review \& editing, Visualization, Supervision}

\affiliation[1]{organization={Computational Biology Unit, University of Bergen},
            addressline={Thorm{\o}hlensgate 55}, 
            city={Bergen},
            postcode={5003}, 
            country={Norway}}

\affiliation[2]{organization={Zuse Institute Berlin},
            addressline={Takustrasse 7}, 
            city={Berlin},
            postcode={14195}, 
            country={Germany}}

\begin{abstract}
The Robust Perron Cluster Analysis (PCCA+) has become a popular spectral clustering algorithm for coarse-graining transition matrices of nearly decomposable Markov chains with transition states. Originally developed for reversible Markov chains, the algorithm only worked for transition matrices with real eigenvalues.
In this paper, we therefore extend the theoretical framework of PCCA+ 
to Markov chains with a complex eigen-decomposition.
We show that by replacing a complex conjugate pair of eigenvectors by their real and imaginary components, a real representation of the same subspace is obtained, which is suitable for the cluster analysis. We show that our approach leads to the same results as 
the generalized PCCA+ (GenPCCA), which replaces 
the complex eigen-decomposition by a conceptually more difficult real Schur decomposition.
We apply the method on non-reversible Markov chains, including circular chains, and demonstrate its efficiency compared to GenPCCA.
The experiments are performed in the Matlab programming language and codes are provided.
\end{abstract}


\maketitle
\section{Introduction}\label{Introduction}

     In order to understand biological processes (e.g., gene regulatory networks \cite{chu2017markov,tse2018rare}, biomolecular dynamics \cite{swope2004describing,noe2009constructing,fersht1995characterizing}), we need to  interpret their dynamics by identifying  structural changes on a network scale. This can be challenging given that biological processes occur on several different time-scales \cite{chodera2014markov}. The available data, however, often represent only the fast time-scale \cite{burke2020biochemical,chodera2014markov}.
     The data describing such fast processes can originate from, for example, mathematical models (e.g., the chemical master equation \cite{tse2018rare}), biomolecular simulations \cite{reuter2018generalized}, or laboratory experiments \cite{fersht1995characterizing}. The data are usually characterized by a huge number of system states, $N$, 
    and stochastic state transitions with short kinetic distances (i.e., on fast time-scales) \cite{pande2010everything}.
     The underlying stochastic process can be described mathematically as a discrete-time Markov chain with a short lag-time and short-lived states (also known as \textit{microstates}) \cite{roblitz2013fuzzy}.  
    
     The system states and transition rates can be summarized into a transition matrix $P$, which allows to calculate the specific probability that the process, in its next  step, switches from one particular state to another \cite{ANDRILLI2010491}.
Upon specific system conditions and assumptions (e.g., thermodynamic equilibrium), and by analyzing the eigendecomposition of the matrix $P$, we can furthermore obtain  additional information about the system dynamics \cite{husic2018markov}.
Such information includes, for example, the identification of metastable sets (also known as \textit{macrostates}). Macrostates represent long-lived system states on the slow time-scale (i.e., with long kinetic distance) and are therefore defined as sub-sets of the state space, in which the system spends a long time before transitioning to the next macrostate \cite{fackeldey2018spectral}. 
By reducing the system from (possibly infinitely) many microstates to a few macrostates, it is easier to  analyze the global process structure and to interpret and understand the system dynamics and the general transition behavior of the Markov chain on long times scales \cite{roblitz2013fuzzy} 

By using spectral clustering methods, one can identify the macrostates through the eigen-decomposition of the matrix $P$ \cite{fackeldey2018spectral}. 
An eminent example of such a method is the Robust Perron Cluster Analysis (PCCA+), which groups microstates into macrostates such that  transitions between macrostates are slow, and transitions within them are fast
\cite{roblitz2013fuzzy}.
The result is a coarse-grained system representation, also known as \textit{Markov State Model (MSM)}, that describes the system behavior on a long time-scale  \cite{chodera2014markov,husic2018markov,reuter2018generalized}. A MSM consists of a finite, discrete state space of size $N$and  a coarse-grained transition probability matrix, $P_c$, which contains the conditional probabilities for transitions between the $n$ ($n<<N$) metastable sub-sets (macrostates) of the system \cite{chodera2014markov}.

As described in \cite{roblitz2013fuzzy}, PCCA+ works for both reversible and non-reversible Markov chains with real dominant eigenvalues, but it does not consider cases with complex eigenvalues. Complex eigenvalues, however, are common for transition matrices of non-reversible Markov chains, and characteristic for circular chains. Therefore, in 2018, the \textit{generalized} PCCA+ method (GenPCCA, sometimes also denoted as G-PCCA) \cite{fackeldey2018spectral,reuter2019generalized}, which employs a conceptually more difficult real Schur-decomposition instead of an eigen-decomposition, was developed.
GenPCCA includes two sub-routines, one for the calculation of the full Schur-decomposition and one for sorting the Schur-vectors. As reported in \cite{brandts2002matlab}, these two sub-routines are computationally ineffective for large eigenvalue problems in Matlab.

The method we present in this article extends the application domain of PCCA+ \cite{roblitz2013fuzzy} to Markov chains with a complex eigen-decomposition. 
We refer to this new approach as cPCCA+, where 'c' stands for 'complex-valued', in order to contrast it with the Schur based GenPCCA and the original PCCA+.

The article is organized in the following way. We start by reviewing the methodological background of  PCCA+ and GenPCCA.
We then present cPCCA+ and show  both theoretically and with illustrative examples that the invariant sub-space condition holds for transition matrices with complex eigenvalues, upon a separation of real and imaginary parts. 
Following, we compare the runtimes of cPCCA+ and GenPCCA for increasing state-space size for two cases, a circular transition matrix and the generator of a stochastic gene-regulatory network. Finally, we discuss the results and present a conclusion. The appendix contains additional figures. A GitHub link to Matlab codes  of  numerical experiments is provided.

\section{Methodological Background}\label{sex:pcca}
Decomposable Markov chains have transition matrices which, upon re-ordering of states according to the partitions, are block-diagonal. The PCCA+ algorithm as a spectral clustering method exploits the fact that the dominant eigenvectors of such matrices, i.e., the eigenvectors corresponding to the multiple eigenvalue one (Perron eigenvalue), are constant on the blocks. In other words, if there are $n_c$ blocks and the matrix $X$ contains the $n_c$ dominant eigenvectors, $X=[X_1,\ldots,X_{n_c}]$, then the rows of $X$ that belong to states in the same block are identical. In fact, each row of $X$ represents one of the $n_c$ corners of an $(n_c-1)$-dimensional simplex (the reduction by one dimension is possible because $X$ contains the constant vector). This simplex structure gets only slightly perturbed if the Markov chain is only nearly decomposable and if transition states occur \citep{roblitz2013fuzzy}. Once this simplex structure is identified, the rows of $X$, i.e., the states of the Markov chain, can be assigned to each of the $n_c$ simplex corners/clusters in terms of fuzzy membership values according to their location within the simplex. For example, the centroid of a simplex would be assigned with membership value $1/n_c$ to each of the $n_c$ clusters. In mathematical terms, finding this simplex structure and the membership values corresponds to finding a linear transformation $A$   

\begin{equation}
\centering
\label{eq:transformation}
\begin{aligned}
 & \chi = X A, \quad~A\in\mathbb{R}^{n_c\times n_c}~\mbox{non-singular}\\
\textrm{s.th.} \quad & \chi_j\geq0~ \forall i\in\{1,...,N\}, j\in\{1,...,n_C\}~\mbox{ (positivity)}\\
  &\sum_{j=1}^{n_c}\chi_{j}(i)=1,~\forall i \in\{1,...,N\} ~\mbox{ (partition of unity)},    \\
\end{aligned}
\end{equation}

where $\chi_j\in\mathbb{R}$ are the \textit{membership vectors}. They assign every state $i\in\{1,...,N\}$ in the state space to a cluster $j\in\{1,...,n_c\}$ with a certain grade of membership \cite{roblitz2013fuzzy,weber2018implications}.

In the following, we assume that the eigenvectors satisfy an \textit{orthonomality relation}
\begin{equation}
    X^{T}D^2X=I
\end{equation}
with identity matrix $I$ and diagonal matrix $D=\text{diag}(w)$, whereby $w$ is some density vector, i.e., $w>0$ and $w^T\boldsymbol{1}=1$. 

\begin{remark}In the implementation of PCCA+ \cite{roblitz2013fuzzy}, $w$ was chosen to be the stationary density, i.e., $w^T P=w^T$. Since the stationary density is very sensitive for nearly uncoupled Markov chains, its computation can be numerically unstable and lead to errors in the orthonormalization of the eigenvectors. The following results, however, are valid for any diagonal matrix $D$, in particular also for $w=(1/N,\ldots,1/N)^T$. 
\end{remark}

As shown in \citep{deuflhard2005}, the problem of finding the right simplex and hence a suitable transformation matrix $A$ does not have a unique solution. Hence, an optimization criterion needs to be defined to make the problem well-posed.
The objective function in PCCA+ is defined as
\begin{equation}
\label{ObjFnc}
    n_c-\text{trace}(D_c^{-2}\chi^TD^2\chi)\rightarrow\text{min},
\end{equation}
whereby $D_c^2=\text{diag}(\chi^T w)$. This optimization results in membership vectors $\chi$ that are as close to characteristic vectors (vectors containing only entries 0 and 1) as possible.

For a given number of clusters $n_c$, the PCCA+ routine consists of two steps.
First it computes an initial guess based on a heuristic, the inner simplex algorithm.
It then proceeds to optimize the objective function using an optimization method, leading to a locally optimal clustering $\chi$. As optimization method, the user can currently choose between the Nelder-Mead simplex method, the Gauss-Newton method, and the Levenberg-Marquardt method.
If the number of clusters is not known in advance, one can maximize the so called \textit{crispness}, $\text{trace}(D_c^{-2}\chi^TD^2\chi)/n_C\in(0,1]$ over $n_c$ to find the optimal number of clusters. In practice one may fall back to the heuristic initial guess for finding the optimal cluster number first before running the whole optimization for that number.
Note that the optimization method for the final cluster number can be chosen differently from the method used to find the optimal cluster number. Typically, the Nelder-Mead method, which is fast but less accurate, is used for finding the optimal cluster number, whereas the slower but more accurate Gauss-Newton method is used for the final optimization. 

In \cite{roblitz2013fuzzy,fackeldey2018spectral}, it is shown that the \textit{invariant sub-space condition} \begin{equation}
    PX=\Lambda X,~\mbox{with}~\Lambda\in\mathbb{C}^{n_{c}\times n_c}
\end{equation}

in combination with the linear transformation $\chi=XA$ and the orthonormality relation  ensures the existence of a coarse-grained propagator matrix that preserves the dominant eigenvalues and hence the time scale of the slow processes. This coarse-grained matrix is defined as 
\begin{eqnarray}
\label{eq:PCCA}
\centering
P_{c}:=(\chi^{T}D^{2}\chi)^{-1} \chi^{T}D^{2}P\chi.
\end{eqnarray}
In fact, it holds
\[P_c=A^{-1}\Lambda A,\]
and hence
\[(P_c^T)^k\chi^T\eta=\chi^T(P^T)^k\eta,\]
which means that the propagagation of the projected density vector (left) commutes with the  projection of the propagated density $\eta$ (right).

Interestingly, any linear transformation of the invariant subspace $X$ preserves both the invariant subspace condition as well as the property of the column vectors to be almost constant on the blocks of a nearly decomposable Markov chain because identical rows in $X$ will also be identical after a linear transformation. 

Since PCCA+ \cite{roblitz2013fuzzy} was  developed for real dominant eigenvectors, it is unable to deal with complex values, which can occur for non-reversible Markov chains. 
In order to maintain the invariant sub-space and to handle  complex eigenvalues, the authors in \cite{fackeldey2018spectralZIB} therefore suggest to work with a real Schur-decomposition of $P$, instead of an eigen-decomposition.
After rearranging the real Schur-matrix $\Lambda$ with the method described in \cite{brandts2002matlab}, such that the $n_c$ Schur-values are close to the Perron root 1, the authors scale the corresponding Schur-vectors by $D^{-1/2}_{\mu}$, where $D_{\mu}$ is a diagonal matrix with an arbitrary density $\mu$ on its diagonal. They show that the scaled Schur vectors satisfy both the invariant sub-space condition and orthogonality relation. GenPCCA then applies the PCCA+ algorithm to the scaled Schur-vectors to solve the optimization problem  (\ref{ObjFnc}) in order to find the transformation matrix $A$. 
Finally, GenPCCA calculates the coarse-grained propagation matrix $P_{c}$ identical to (\ref{eq:PCCA}) as\begin{eqnarray}
\label{eq:GPCCA}
\centering
P_{c}=(\chi^{T}D_{\mu}^{2}\chi)^{-1} \chi^{T}D_{\mu}^{2}P\chi.
\end{eqnarray}

In the next section, we present our new approach which employs an eigen-decomposition.
\section{Complex-valued PCCA+ (cPCCA+)}
\label{FPCCAplus}
As theoretical basis for cPCCA+, we present Theorem~\ref{thm1}, which shows how to construct the real subspace from the complex eigendecomposition.
\begin{theorem}{(Invariant sub-space condition for complex eigenvectors)}
\label{thm1}
Replacing a complex conjugate pair of eigenvectors in the eigenvector matrix by the real and imaginary parts of these eigenvectors leads to a matrix that spans the same invariant subspace as the original eigenvector matrix.
\end{theorem}

\begin{pot1}
Let $X\in\mathbb{C}^{N\times~n_{c}}$ be the matrix of eigenvectors, and assume that columns $k$ and $k+1$ contain a complex conjugate pair of eigenvectors $x_k=a_k+b_k\i$ and $\bar {x}_k=a_k-b_k\i$ with $a_k,b_k\in\mathbb{R}^N$ and imaginary unit $\i$. By replacing the columns $x_k$ and $\bar{x}_k$ by $a_k$ and $b_k$, we obtain a new matrix $\tilde{X}$. 

We now show that $\tilde{X}$ spans the same subspace as X by expressing it as a linear combination of X,
\[\tilde{X}=X\cdot C,\]
where $C\in\mathbb{R}^{n_c\times n_c}$ only acts on columns $k$ and $k+1$:
\[C=\begin{pmatrix}1&0&\ldots&&&&&\\
0&\ddots&&&&&&\\
\vdots&&1&&&&&\\
&&&c_1&c_3&&&\\
&&&c_2&c_4&&&\\
&&&&&1&&\vdots\\
&&&&&&\ddots&0\\
&&&&&\ldots&0&1
\end{pmatrix}\]
In terms of columns $k$ and $k+1$, the system of equations $\tilde{X}=X\cdot~C$ reads
\begin{eqnarray}
\centering
\left[
\begin{array}{cc}
  a_k & b_k 
\end{array}
\right]=
\left[
\begin{array}{cc}
  a_k+b_k\i & a_k-b_k\i  
\end{array}
\right]\times 
\begin{pmatrix}
    c_1 & c_3 \\ c_2 & c_4 
\end{pmatrix}
\end{eqnarray}
with the unique solution 
\begin{align}
    \begin{pmatrix}
    c_1 & c_3 \\ c_2 & c_4 
    \end{pmatrix}
    = 
    \frac{1}{2}\begin{pmatrix}
    1 & -i \\ 1 & i 
    \end{pmatrix}.
\end{align}
Hence, the $2\times 2$ matrix formed by $c_1$ to $c_4$ and consequently $C$ is regular. The linear combination $X\cdot C$ is thus an invariant sub-space of $P$ again.
\qed
\end{pot1}

Since both cPCCA+ and GenPCCA work on the same subspace, we can conclude that they are equivalent in the following sense:

\begin{corollary}\label{Col1}
If $P$ admits an eigen-decomposition, cPCCA+ and GenPCCA  obtain the same solutions to the optimization problem.
\end{corollary}
\begin{pf}
Both algorithms solve the original PCCA+ problem (\ref{eq:transformation}-\ref{ObjFnc})
with different bases $X_1, X_2$ of the same subspace $\text{span}(X_1) = \text{span}(X_2)$. Thus there exists a regular matrix $B$ s.t. $X_1 = X_2  B$. Then any solution to the objective function of the form
$\chi = X_1  A_1$ is also a solution of the form $\chi = X_2 B A_1 = X_2  A_2$ with $A_2 = BA_1$ and vice versa.
\qed
\end{pf}

Note that in practice we use only local optimization and hence the different representations and resulting initial guesses may lead to different local minima which, however, are also local minima to the respective other problem.

In the following, we will consider examples for the case of (i) a nearly uncoupled Markov chain, and (ii) a Markov chain with circular transition pattern, and argue in both cases for the fulfillment of the invariant subspace condition.

\subsection{Non-reversible, nearly uncoupled Markov chains}

 In Example~\ref{exp:1}, we consider the case of a small,  non-reversible system with three metastable states and show that cPCCA+ clusters the states  accordingly. 

\begin{figure}[h!]
	\centering
		\includegraphics[scale=0.28]{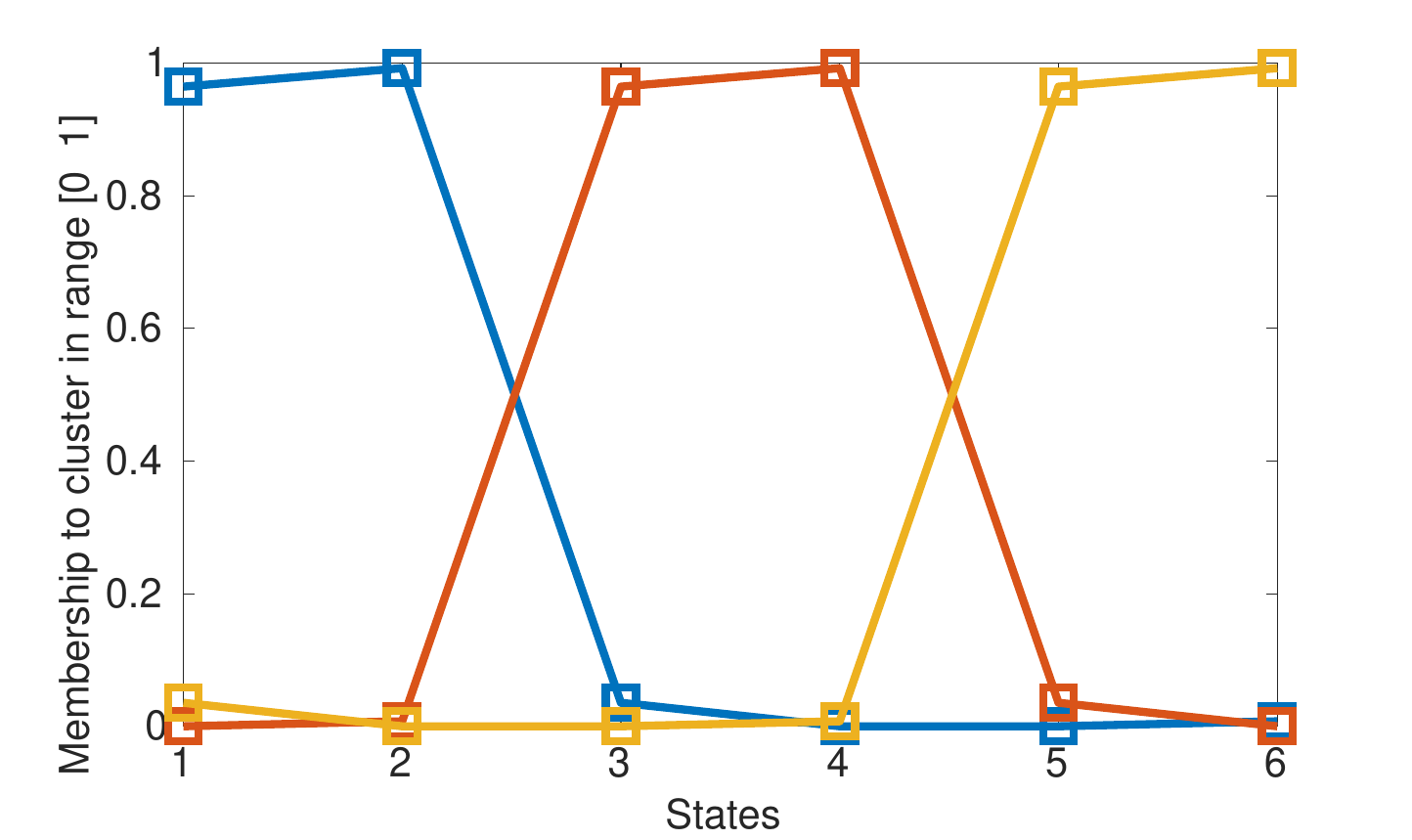}
	  \caption{Membership functions $\chi$ for Example 1: The functions show that the six system states are clustered into three distinct groups, each containing two states. } \label{Fig:Exp1NonRev}
\end{figure}

\begin{Exp}\label{exp:1}
\begin{figure}
    \centering
    \includegraphics[height=7cm]{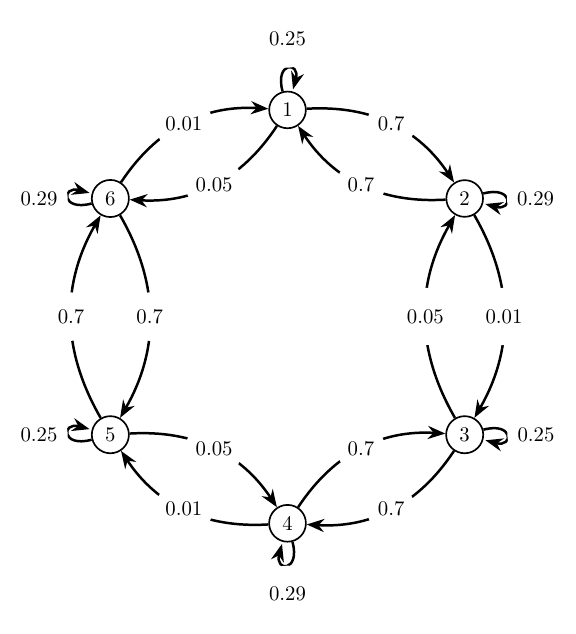}
    \caption{Weighted directed graph for the Markov chain in Example \ref{exp:1}.}
    \label{Fig:Exp1-graph}
\end{figure}

Let $P\in\mathbb{R}^{6\times~6}$ be the transition matrix for the Markov chain in Fig.~\ref{Fig:Exp1-graph}:
\begin{eqnarray}
\centering
P=\left[
\begin{array}{cccccc}
  0.25 & 0.7 & 0 & 0 & 0&0.05 \\
   0.70 & 0.29 & 0.01 & 0 & 0&0 \\
    0& 0.05 & 0.25 & 0.70 & 0&0 \\
    0& 0& 0.70 & 0.29 & 0.01&0 \\
    0& 0& 0 & 0.05 & 0.25&0.70 \\
    0.01& 0& 0 & 0 & 0.70&0.29 
\end{array}
\right]
\label{mat:nonRev}
\end{eqnarray}
This matrix has three eigenvalues close to one:  $\lambda_1=1.0000, \lambda_2=0.9557+0.0177\i, \lambda_3=0.9557-0.0177\i$.
Figure~\ref{Fig:Exp1NonRev} shows the fuzzy clustering of the six states into three clusters according to the membership functions $\chi$ computed by cPCCA+. 
Each of the three colors (blue, red and yellow) represents a different cluster. The system states with values of $\chi$ near 1, are those that  strongly belong to a specific cluster. The resulting coarse-grained transition matrix $P_{c}\in\mathbb{R}^{3~\times~3}$ in (\ref{eq:CaseNR}) has diagonal-entries close to unity, which indicates three strong metastable clusters.
\begin{eqnarray}
\centering
P_c=\left[
\begin{array}{ccc}
  0.9705 & 0.0046 & 0.0250 \\
  0.0250 & 0.9705 & 0.0046\\
  0.0046 & 0.0250 & 0.9705
\end{array}
\right]
\label{eq:CaseNR}
\end{eqnarray}
\end{Exp}

This example illustrates that cPCCA+ can deal with non-reversible stochastic matrices with complex dominant eigenvalues.

\subsection{Circular Markov chains}
In this sub-section, we make use of the unique structure of circular matrices.
In Theorem \ref{thm2}, we show that 
the dominant (complex) eigenvectors of circular stochastic matrices are piece-wise constant on the blocks. 

\begin{theorem}{(Eigenvector structure for circular transition matrices)}
\label{thm2}
Let $P\in\mathbb{R}^{(N\times~n)\times(N\times~n)}$ be a row-stochastic, circular matrix with the following structure,
\begin{eqnarray}
\centering
P=\left[
\begin{array}{cccc}
  0 & X_{1} &\cdots & 0 \\
   \vdots & \ddots & \ddots& \vdots \\
   0 & \cdots & 0 &  X_{N-1}\\
  X_{N} & 0 &\cdots & 0
\end{array}
\right],
\end{eqnarray}
where $X_k\in\mathbb{R}^{n\times~n},\, k=1,\ldots,N$ are row-stochastic sub-matrices.
Then the matrix $P$ has $N$ eigenvalues $\lambda_1,...,\lambda_{N}\in\mathbb{C}$ uniformly distributed on the unit circle, 
\[\lambda_{k}=\exp((k-1)~\theta\i)~\mbox{with}~\theta=\frac{2\pi}{N},\quad k=1,...,N.\]

The corresponding eigenvectors $V_k$, i.e., those satisfying $P V_{k}=\lambda_{k} V_{k}$ for $k=1,...,N$, are piece-wise constant on each block $X_k$.
\end{theorem}

\begin{pot2}
The eigenvectors $V_{k}$ corresponding to eigenvalues $\lambda_{k}$ can be written in a general form as
\begin{eqnarray}
\centering
\label{Eq:EigV}
V_{k}^{T}=[\underbrace{\lambda_{k}^1,\cdots,\lambda_{k}^1}_{n},
\underbrace{\lambda_{k}^2,\cdots,\lambda_{k}^2}_{n},
\underbrace{\lambda_{k}^3,\cdots,\lambda_{k}^3}_{n},\cdots,
\underbrace{\lambda_{k}^{N-1},\cdots,\lambda_{k}^{N-1}}_{n},
\underbrace{\lambda_{k}^N,\cdots,\lambda_{k}^N}_{n}]^T.
\end{eqnarray}

Noting that $\lambda_k\cdot\lambda_k^N=\lambda_k^{N+1}=\lambda_k$ and that the matrix $P$ shifts the process in clockwise direction to the next block, it holds
\begin{eqnarray}
\centering
P\cdot V_{k}= [\underbrace{\lambda_{k}^2,\cdots,\lambda_{k}^2}_{n},
\underbrace{\lambda_{k}^3,\cdots,\lambda_{k}^3}_{n},
\underbrace{\lambda_{k}^4,\cdots,\lambda_{k}^4}_{n},\cdots,
\underbrace{\lambda_{k}^N,\cdots,\lambda_{k}^N}_{n},\underbrace{\lambda_{k}^1,\cdots,\lambda_{k}^1}_{n}]^{T}
= \lambda_{k}\cdot V_{k}.
\end{eqnarray}
This shows that the eigenvectors corresponding to eigenvalues on the unit circle are piecewise constant on the blocks. \qed
\end{pot2}

According to Theorem \ref{thm1}, separating the real and imaginary parts of eigenvectors $V_k$ is just a linear transformation which maintains both the invariant subspace condition and the piecewise constant block structure. Hence, cPCCA+ can be applied successfully to circular matrix structures.  

Next, we demonstrate that cPCCA+ also works for a non-reversible transition matrix $P\in\mathbb{R}^{9\times 9}$ where the circular transition pattern is slightly perturbed. The matrix structure in Example~\ref{exp:2} is taken from \cite{fackeldey2018spectral} (see Example 3.1 therein).

\begin{Exp} 
\label{exp:2}
We consider a stochastic matrix $P\in\mathbb{R}^{9\times 9}$,
\begin{eqnarray}
\centering
P=\left[
\begin{array}{cccc}
  X & Y & \mathbf{0} \\\nonumber
    \mathbf{0}&  X & Y\\\nonumber
  Y & \mathbf{0} & X\nonumber
\end{array}
\right],
\end{eqnarray}
consisting of identical diagonal blocks $X\in\mathbb{R}^{3\times 3}$ with the structure
\begin{eqnarray}
\centering
X=\left[
\begin{array}{ccc}
  0 & x & 0 \\
   0 &  0 &  1.0\\
  1.0 & 0 & 0
\end{array}
\right],
\label{eq:Exp1X}
\end{eqnarray}
 and identical off-diagonal blocks $Y\in\mathbb{R}^{3\times 3}$ with the following pattern,
\begin{eqnarray}
\centering
Y=\left[
\begin{array}{ccc}
 y & 0 & 0 \\\nonumber
   0 &  0 &  0\\\nonumber
  0 & 0 & 0 \nonumber
\end{array}
\right].
\end{eqnarray}

The $\mathbf{0}\in\mathbb{R}^{3\times 3}$ represent zero block matrices.

The matrix $P$ contains one block-diagonal pattern, formed by the $X$ blocks in $P$, and additionally two types of circular patterns, one within each $X$ block and one formed by the $Y$ blocks in $P$. Whether one identifies the block-diagonal or the circular pattern depends on the values of ${x}\in\mathbb{R}$ and ${y}\in\mathbb{R}$, as well as on the eigenvalues that are considered as dominant (either the ones with largest magnitude or those with largest real part). In the following, we analyze the four different cases (i)-(iv), as presented in Table~\ref{Tab:1}.
\begin{table}[H]
\begin{tabular}{ p{4cm}||p{3cm}|p{3cm}  }
 \hline
& \multicolumn{2}{c}{Dominant eigenvalue type:} \\
\centering
 Matrix entries for $x$ and $y$:    &largest magnitude& largest real part  \\
     \hline\hline
     $x=0.9,\, y=0.1$& Case (i)& Case (ii)\\
     $x=0.1,\, y=0.9$& Case (iii)& Case (iv)\\
     \hline
\end{tabular}
\caption{The four cases of matrix patterns in Example 2}
\label{Tab:1}
\end{table}
We examine the eigenvalues, the membership functions $\chi$, and the coarse-grained matrices $P_c$ computed by cPCCA+ in each of the four cases.

{\bf Case (i).} The three eigenvalues with largest magnitude are $\lambda_{1}=1.0000+0\i,~\lambda_{2}=-0.5000 + 0.8660\i,~ \lambda_{3}=-0.5000-0.8660\i$.
Figure \ref{Fig:Exp1Case1} shows the clustering of the nine states into three clusters based on the $\chi$ membership functions. The three clusters are characterized by sharing the same transition pattern: cluster 1 (blue -  states 1, 6, 8) transitions to cluster 2 (red - states 2, 4, 9), cluster 2 transitions to cluster 3 (yellow - states 3, 5, 7), and cluster 3 transitions back to cluster 1. That means the symmetry in the transition graph is perfectly recovered.  
The resulting coarse-grained transition matrix reads
\begin{eqnarray}
\centering
P_c=\left[
\begin{array}{ccc}
  0 &1 & 0 \\
  0 & 0&{1}\\
  {1} & 0 &0
\end{array}
\right].
\label{eq:Case1}
\end{eqnarray}

{\bf Case (ii).} If we instead solve for the eigenvalues with largest (real) part, we obtain  $\lambda_{1}=1.0000+0\i,\,\lambda_{2}=0.9483 + 0.0279\i,\,
\lambda_{3}= 0.9483 - 0.0279\i$, indicating that the Markov chain represents a nearly-uncoupled Markov chain.
Figure \ref{Fig:Exp1Case1} shows the clustering of the nine states into three clusters based on the $\chi$ membership functions. The three clusters reflect the block diagonal structure that is visible in $P$: cluster 1 includes states 1, 2, 3 (blue), cluster 2 includes states 4, 5, 6 (red), and cluster 3 includes states 7, 8, 9 (yellow).
The resulting coarse-grained transition matrix has the form
\begin{eqnarray}
\centering
P_c=\left[
\begin{array}{ccc}
{0.9655}  &  0.0333  &   0.0012 \\
 0.0012&   {0.9655}  & 0.0333 \\
0.0333  &   0.0012 &   {0.9655}
\end{array}
\right],
\label{eq:Case2}
\end{eqnarray}
which reflects the decoupling of the chain into three nearly decoupled sub-chains and the perfect symmetry between the clusters. For $x=1$ and $y=0$, the decoupling would be complete and $P_c$ would be the identity matrix with zero transition probabilities between the clusters.

\begin{figure}[h!]
\begin{tabular}{cc}
  	\includegraphics[height=4cm]{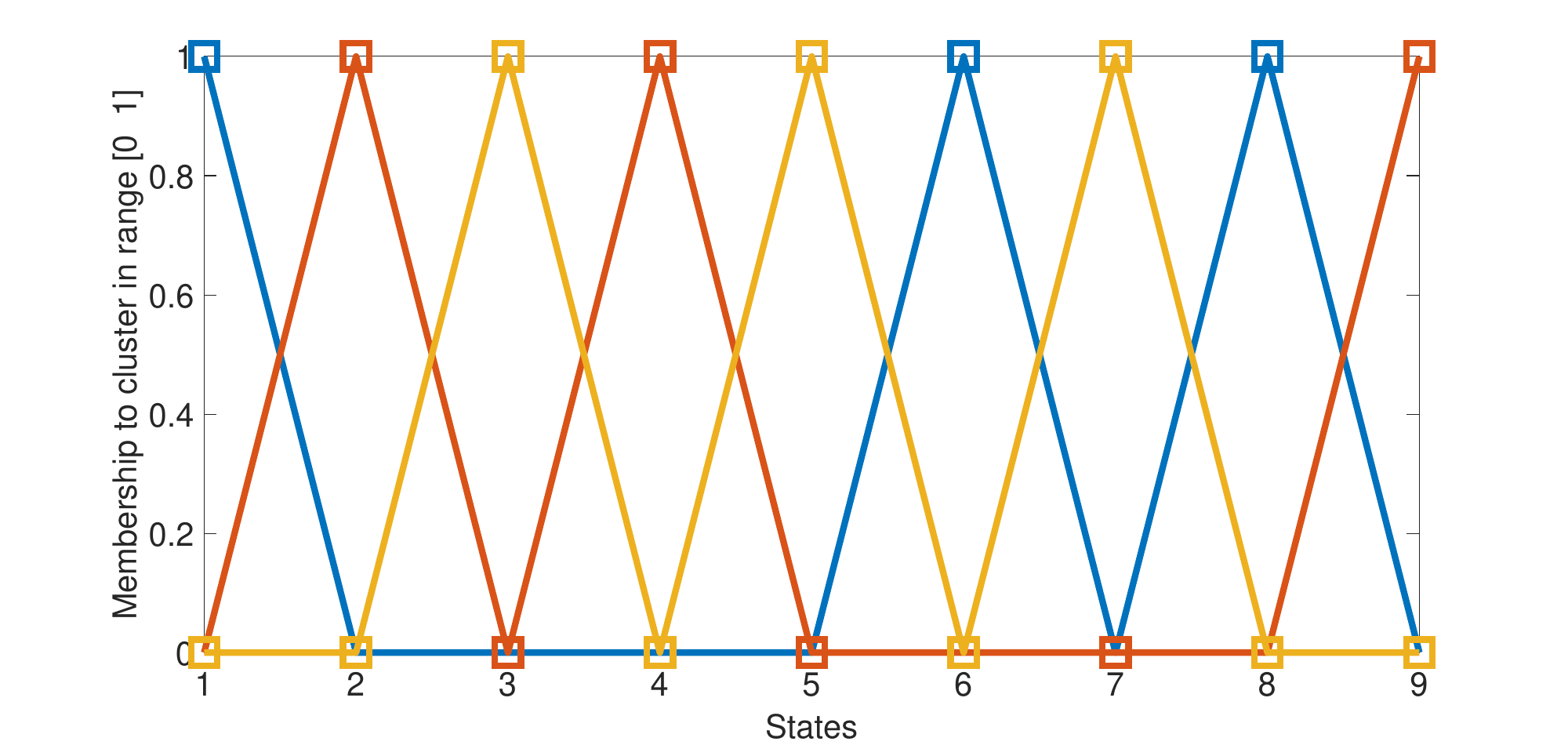} & 	\includegraphics[height=4cm]{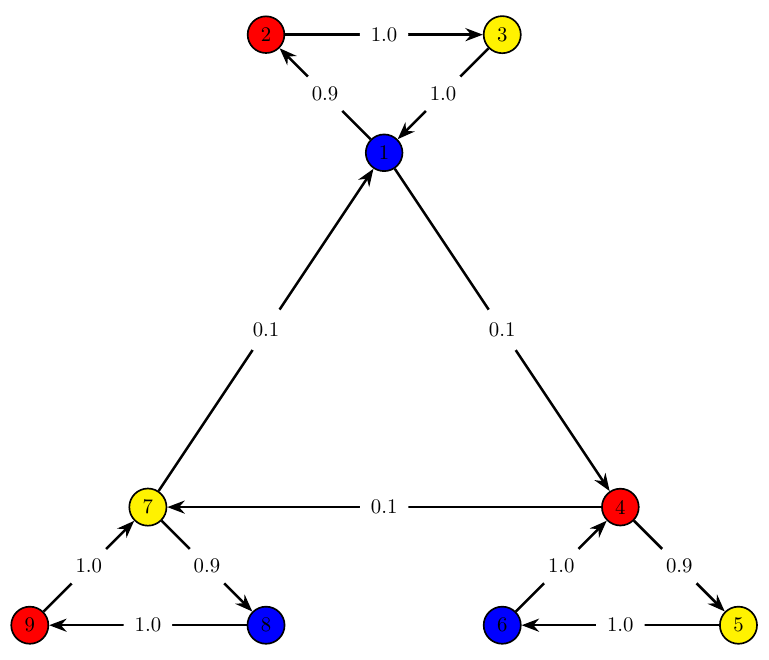} 	 \\
  	  {\bf Case (i).} $x=0.9,\,y=0.1$, \textit{largest magnitude} eigenvalue  &  Three circulating clusters\\[0.5cm]
  	 \includegraphics[height=4cm]{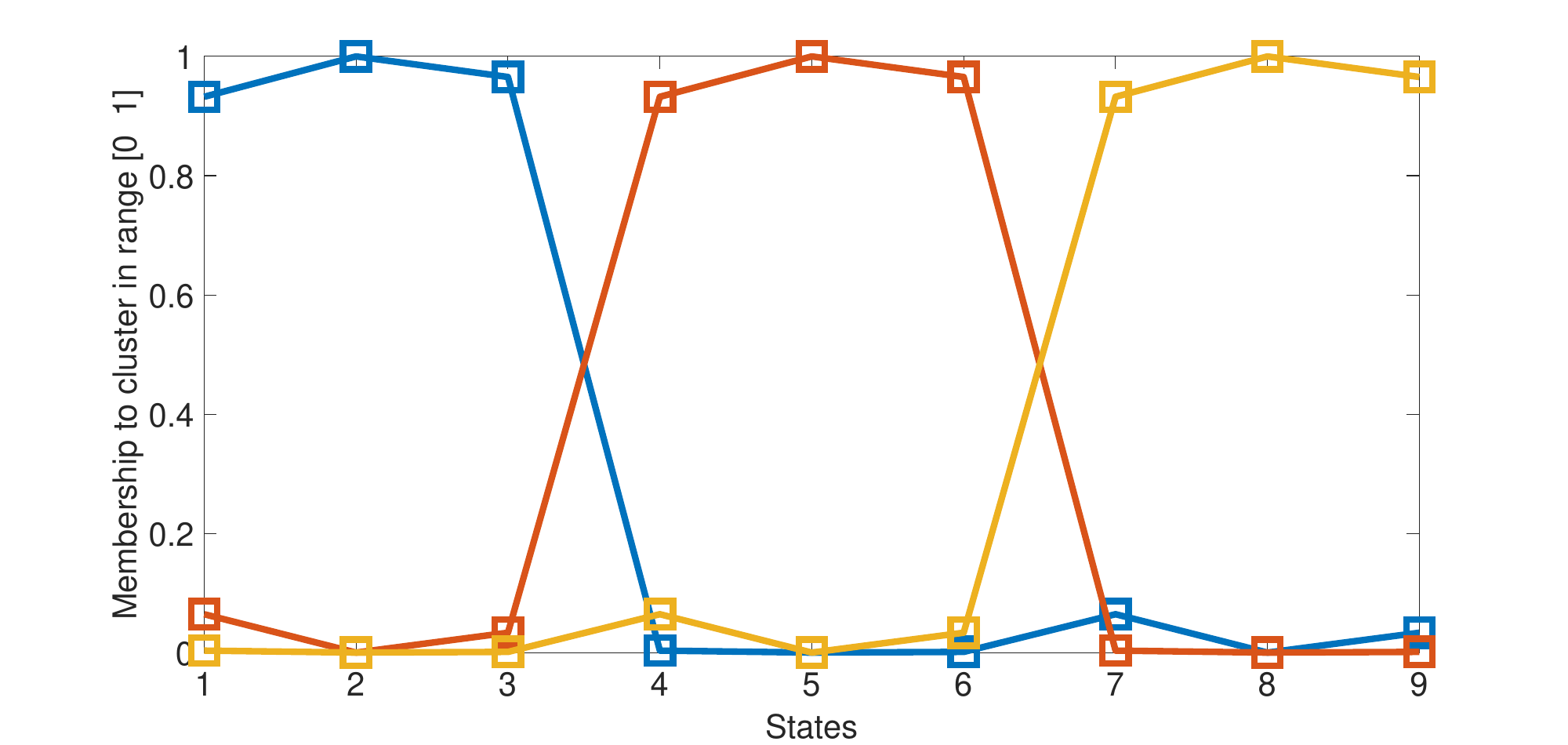} & \includegraphics[height=4cm]{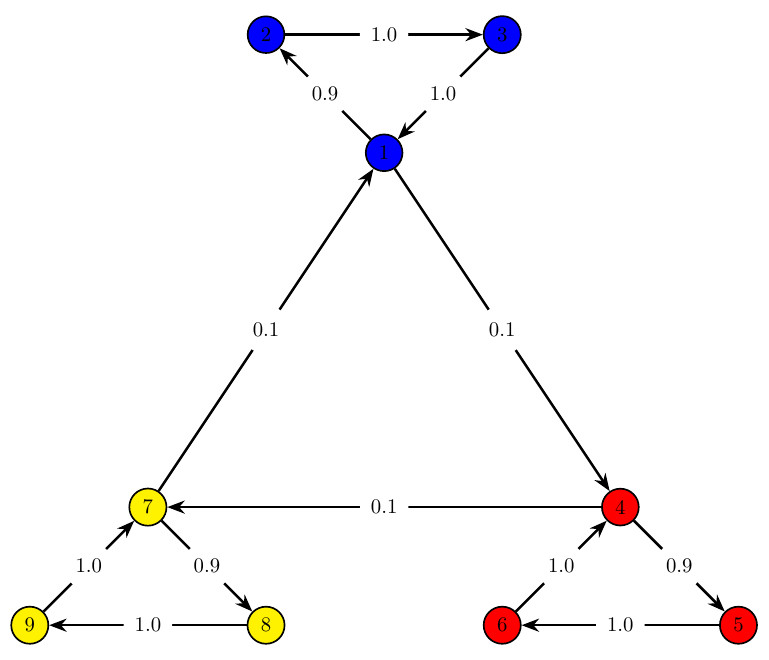}\\
  	  {\bf Case (ii).}  $x=0.9,\,y=0.1$, \textit{largest (real)} eigenvalue  &   Three block-group clusters\\[0.5cm]
  \includegraphics[height=4cm]{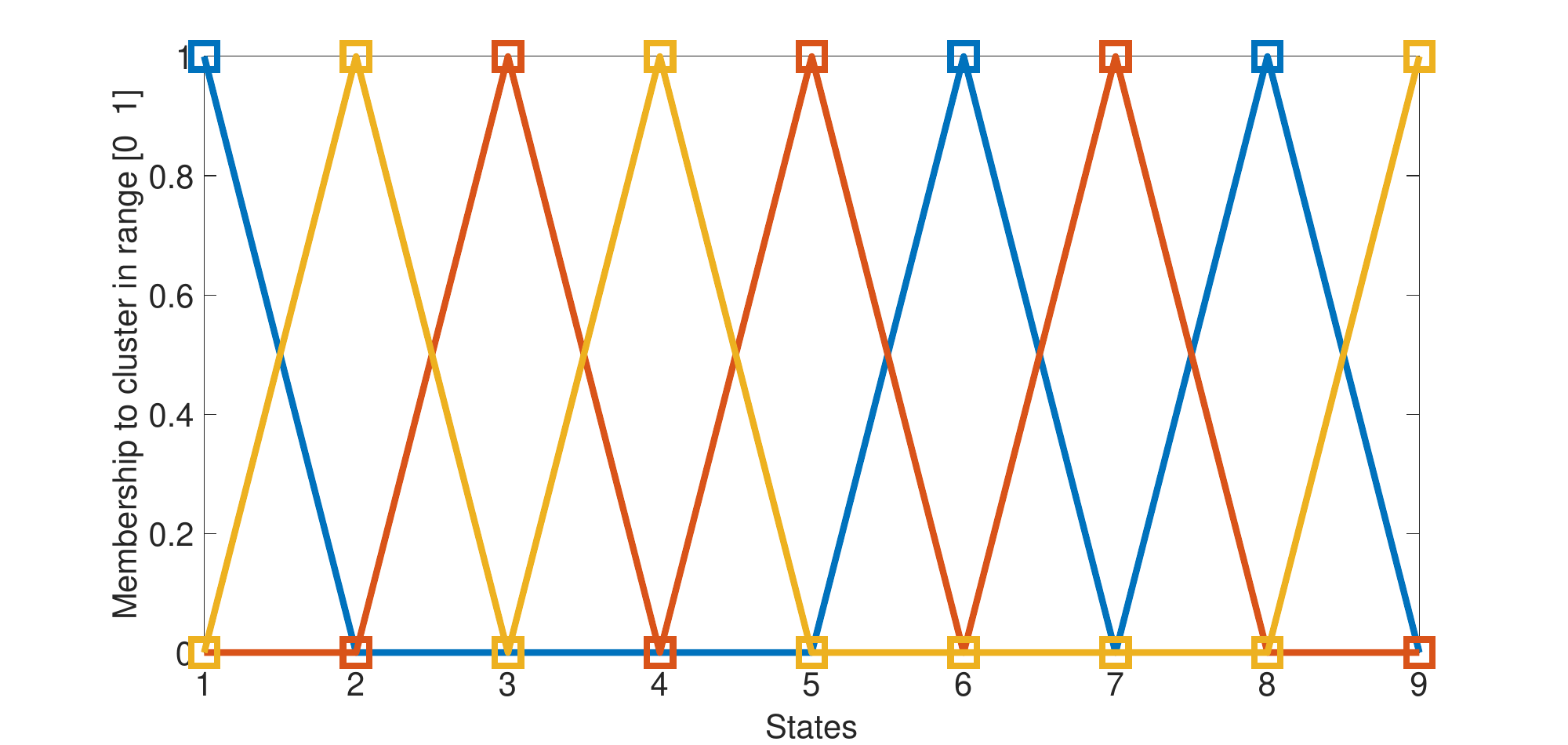} 
  &\includegraphics[height=4cm]{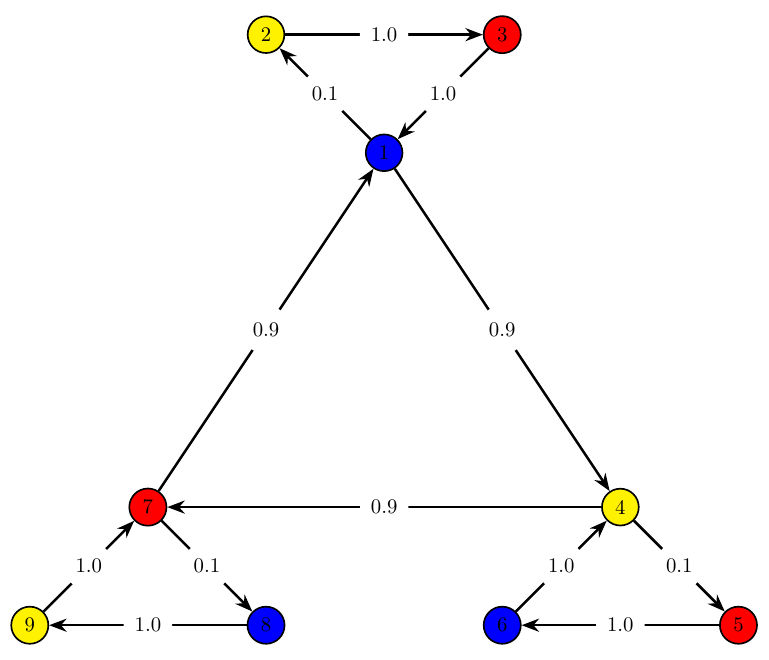} \\
	   {\bf Case (iii).}  $x=0.1,\,y=0.9$, \textit{largest magnitude} eigenvalue &  Three circulating clusters \\[0.5cm]
	     \includegraphics[height=4cm]{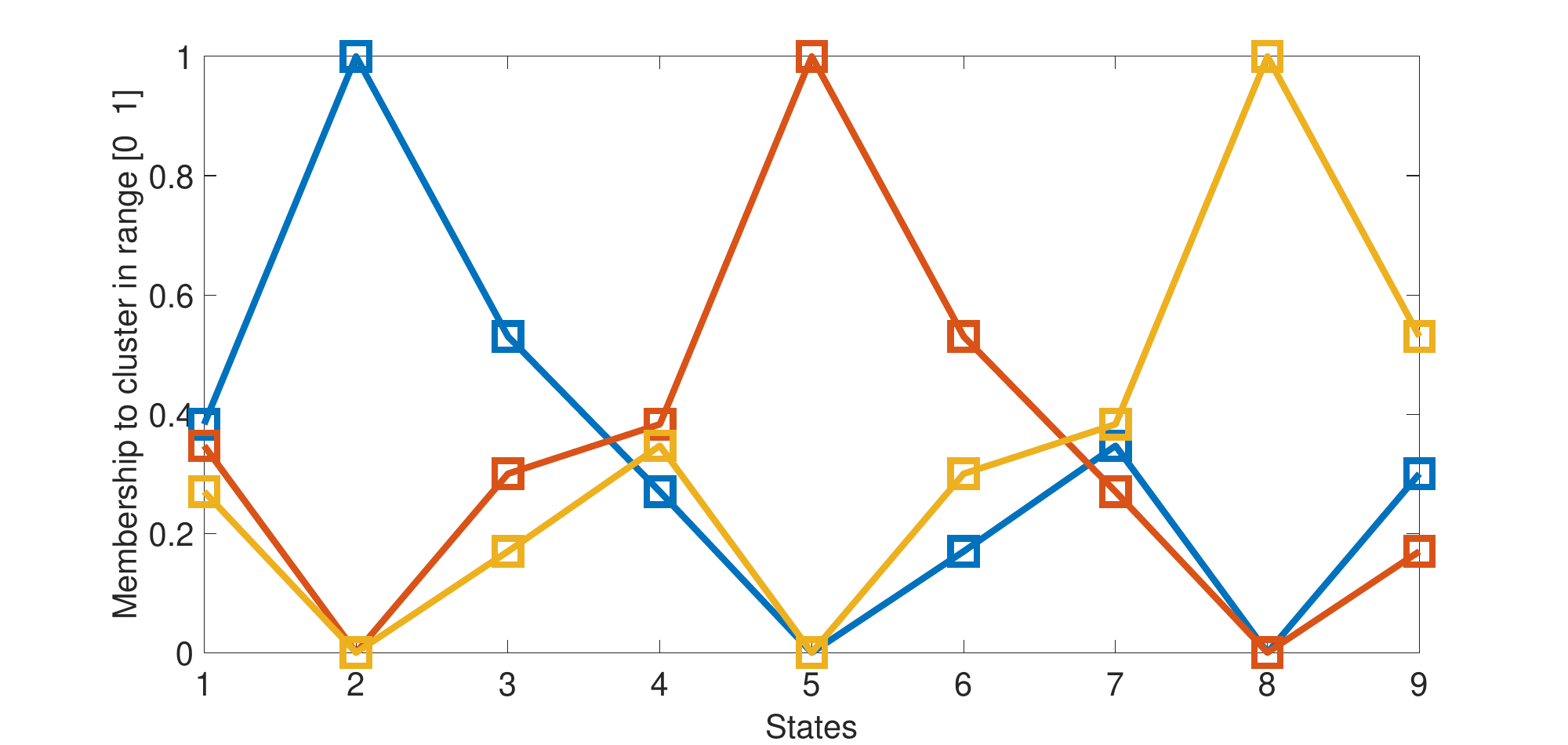} 
  &\includegraphics[height=4cm]{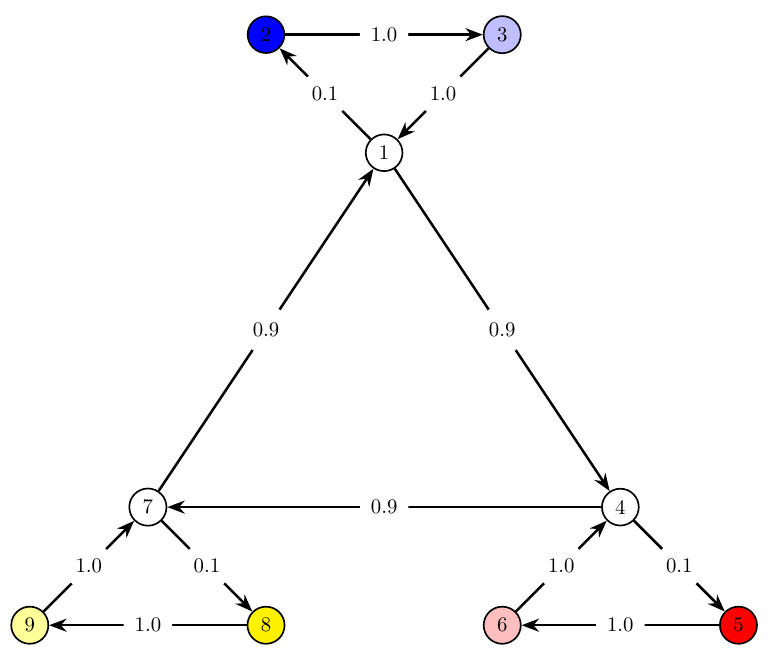} \\
	   {\bf Case (iv).}  $x=0.1,\,y=0.9$, \textit{largest (real)} eigenvalue &  No multistability, hence no clear clustering
\end{tabular}
\caption{Example \ref{exp:2}: Membership functions $\chi$ (left column) and transition networks (right column) }
	\label{Fig:Exp1Case1} 
\end{figure}

{\bf Case (iii).} For $x=0.1$ and $y=0.9$, i.e., an increased coupling between the sub-chains represented by $X$, the three eigenvalues with largest magnitude are (upon permutation) $\lambda_{1}=1.0000+0\i,~\lambda_{2}=-0.5000 + 0.8660\i,~ \lambda_{3}=-0.5000-0.8660\i$, indicating that there is a circular transition pattern. Figure \ref{Fig:Exp1Case1} shows that cPCCA+ identifies the same circular transition pattern as in case (i), except that clusters 2 and 3 are permuted\footnote{The clustering delivered by PCCA+ is unique up to a permutation of membership vectors.}. That means, cluster 1 (blue - states 1,6,8) transitions into cluster 3 (yellow - states 2,4,9), cluster 3 transitions into cluster 2 (red - states 3,5,7), and cluster 2 transitions back into cluster 1:
\begin{eqnarray}
\centering
P_c=\left[
\begin{array}{ccc}
  0 & 0 & {1} \\
   {1} & 0& 0\\
  0 &  {1} &0
\end{array}
\right].
\label{eq:Case3}
\end{eqnarray}

{\bf Case (iv).}
Applying cPCCA+ to Case (ii) specifying largest (real) eigenvalue, results in only one eigenvalue close to 1 ($\lambda_{2,3}=0.2954\pm 0.1128\i$), i.e., there is no multi-stability. If we nevertheless enforce a clustering into three groups, then states 2, 5, and 8 form the centers of these three clusters, because the transition probabilities between these three states are very small. States 1, 4, and 7 are assigned to the three clusters with almost equal membership, indicating that they are transition states between the three clusters. Each of the states 3, 6, and 9 is assigned to one of the three clusters with a membership value $>0.5$, which shows that they are stronger associated with the clusters than the states 1, 4, and 7.  
The coarse-grained transition matrix (for cluster order blue - red - yellow) has the form
\begin{eqnarray}
\centering
P_c=\left[
\begin{array}{ccc}
  {0.5303}  &  0.3000&    0.1697\\
    0.1697   & {0.5303} &   0.3000\\
    0.3000   & 0.1697&   {0.5303}
\end{array}
\right],
\label{eq:Case4}
\end{eqnarray}
which underlines that the three clusters are only weakly decoupled. 
\end{Exp}

In the following, we compare the numerical runtimes and clustering results of cPCCA+ and GenPCCA for two larger matrix examples in Matlab.

\section{Comparison of numerical runtimes}
We consider the cases of 
(1) a circular transition matrix (taken and adapted from \cite{fackeldey2018spectral}; see Example 3.2 therein), and (2) the infinitesimal generator of a stochastic, non-reversible gene-regulatory network, generated from the model in \cite{frank2021bifurcation}. 
For both examples, we  compare the computational elapse time (ET) for varying state-space sizes as well as the normed differences between the coarse-grained transition matrices $P_c$ obtained from cPCCA+ and GenPCCA, respectively. All computation were done in Matlab (version '9.7.0.1261785 (R2019b) Update 3'). The ET was calculated with the \textit{tic}-\textit{toc} commands and the normed differences were determined with the \textit{norm(difference,p)} function with  $p\in\{ 1,2,\infty\}$.

Next, we present the generation process of the example matrices and the results of the comparison for the two cases.

\begin{enumerate}
\item {\bf Circular transition patterns with and without perturbations.}
In this case, we analyze circular matrices  $P$ with varying state-space size $N$ for $N=30,60,90,120$. 
To construct the matrices $P$, we proceed similar to Example 3.2 in \cite{fackeldey2018spectral}: We start by creating three n-by-n random matrices A1, A2, and A3 using the Matlab-routine “rand(n,n)”, and one n-by-n zero matrix Z, where $n$ takes the values 10, 20, 30 and 40. A circular matrix structure is attained by defining another matrix C, via C:=[[Z,A1,Z];[Z,Z,A2];[A3,Z,Z]], which varies in state-space size as specified above. Upon re-scaling, matrix $C$ becomes stochastic and represents the non-perturbed circular matrices $P$ to be analyzed.

In order to create perturbed stochastic circular matrices $P$, a random matrix containing entries between 0 and 0.1 is added to matrix C before it is re-scaled.

To account for variability in the random generation of the circular matrices and to estimate the mean and standard deviation of ETs and clustering results, we generate, for each matrix size $N$ and each case (perturbed and non-perturbed), five different matrices $P$ (according to the steps described above).

The results show that both in the non-perturbed case (Fig.~\ref{Fig:ETCircularUnP}) and in the perturbed case (Fig.~\ref{Fig:ETCircularPert}), cPCCA+ and GenPCCA are fast, but that cPCCA+ is consistently quicker.

\begin{figure}[h!]
	\centering
		\includegraphics[scale=0.25]{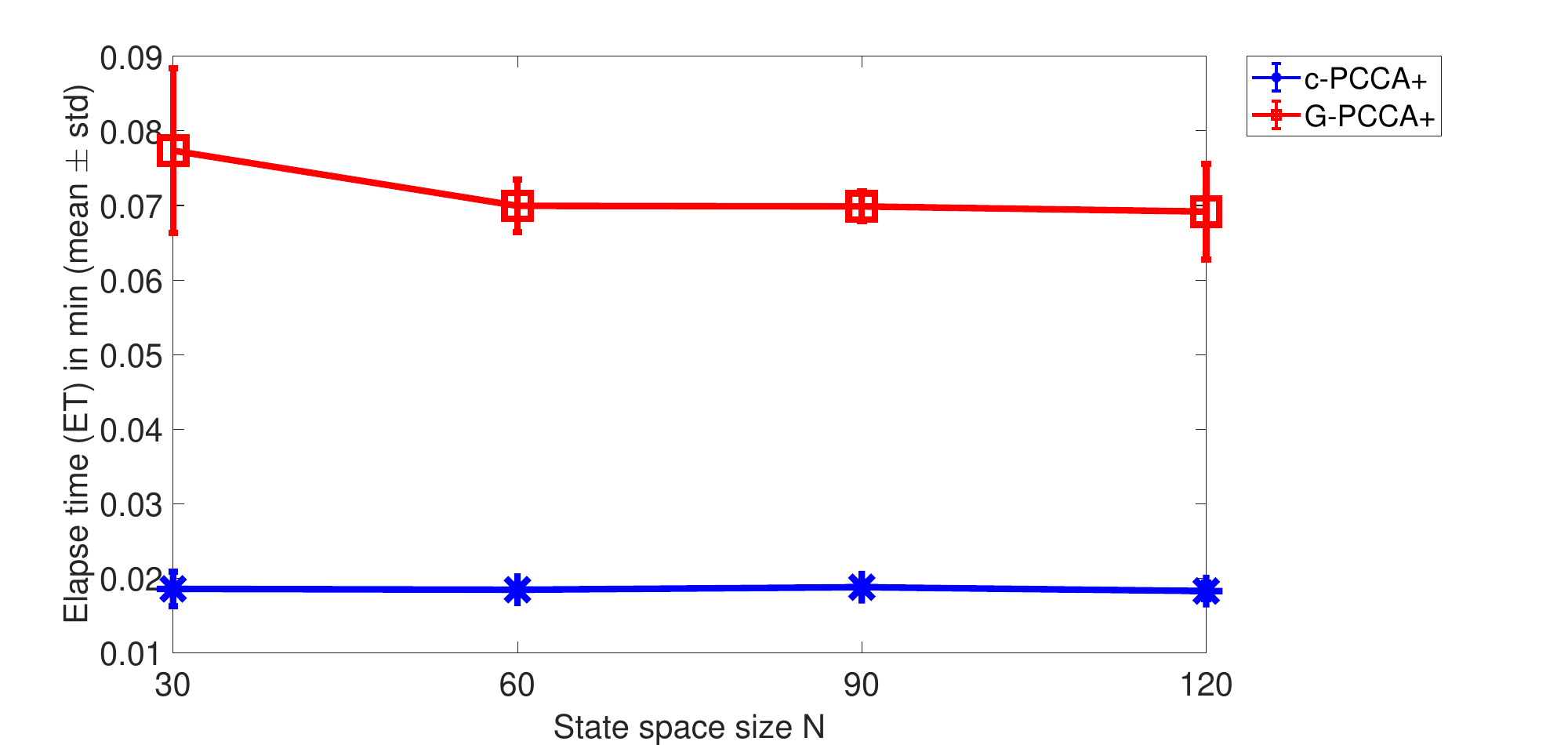}
	  \caption{Elapse time (ET) of cPCCA+ and GenPCCA for non-perturbed circular matrix structure. The figure shows an error bar plot with the mean and standard deviation of five ETs for each state space size $N$. }\label{Fig:ETCircularUnP}
\end{figure}

 We also see in Figures \ref{Fig:CircularDiff} (non-perturbed case) and \ref{Fig:PertCircDiff} (perturbed case) that the differences in the coarse-grained transition matrices are small, although it seems that the difference in the perturbed case increases with matrix size (Figure \ref{Fig:PertCircDiff}).
 
\begin{figure}[h!]
	\centering
		\includegraphics[scale=0.25]{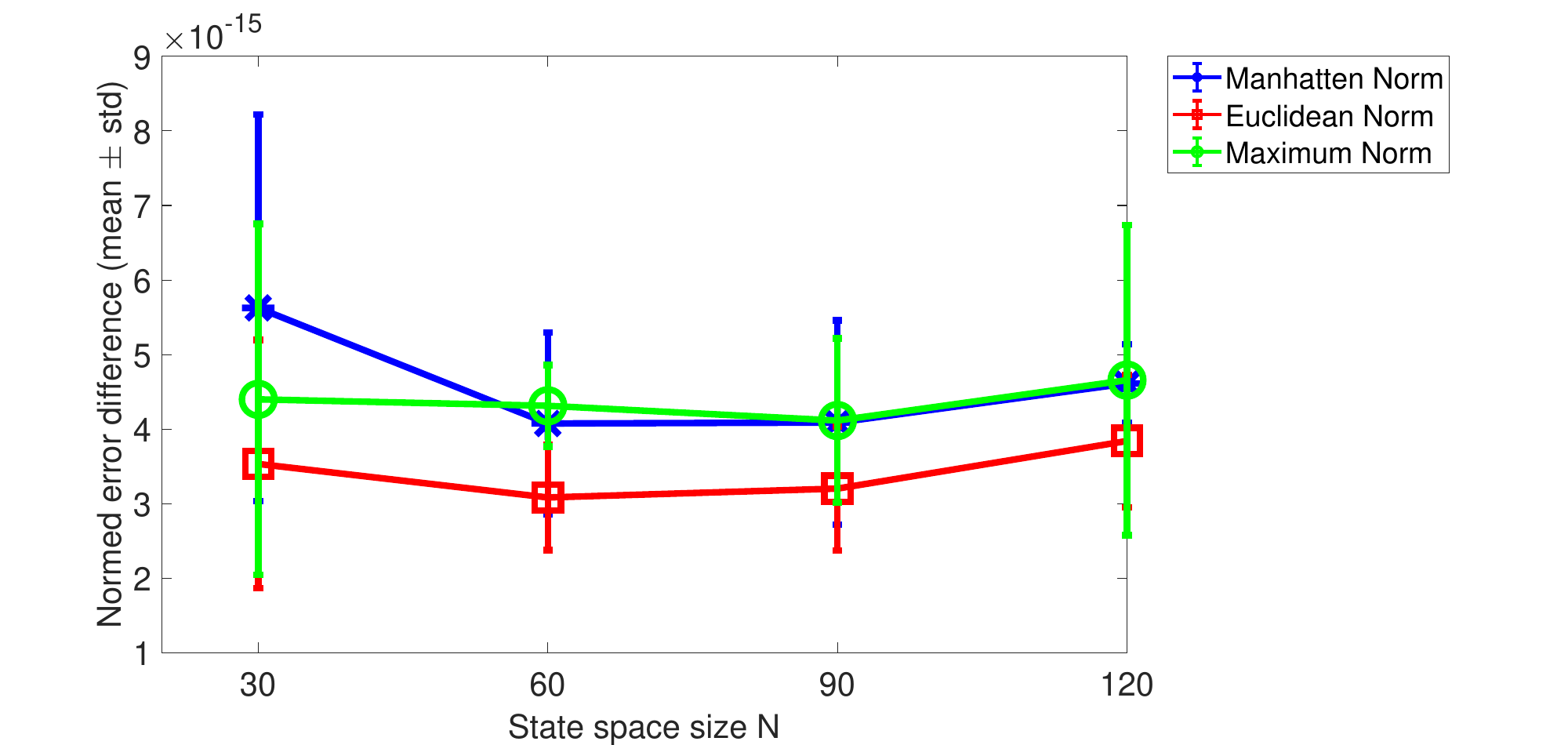}
	  \caption{Normed error differences in coarse-grained transition matrices $P_c$ for the non-perturbed circular matrices. The error bars are based on the mean and standard deviation of five normed differences in $P_c$ between the cPCCA+ and the GenPCCA algorithm for each state space size $N$. } \label{Fig:CircularDiff}
\end{figure}

\item {\bf Stochastic gene-regulatory network.}
We transformed the 2-dimensional deterministic macrophage polarization model described in \cite{frank2021bifurcation} into a stochastic gene-regulatory network whose dynamic is described by the Chemical Master Equation (CME).
We analyze the system on a finite state space of size $N^2$ for $N\in\{20,40,60,80,100\}$. 
For the considered parameter set, the system shows tri-stability.

Making use of the calculated ET for the cPCCA+ and GenPCCA, we determine the procentual difference in ETs between these two methods, to which we fit a quadratic polynomial curve ($f(x)=a~x^2+b~x+c$). This approach allows us to quantify a functional relationship between the runtimes of both methods.
In addition, we aim to identify the  GenPCCA sub-routine in which the algorithm spends most time. To this end, we determine ETs for different subroutines in the GenPCCA algorithm. These sub-routines are the minChi optimization routine, optimization procedures with methods 'nelder-mead' (optimization loop 1) and 'gauss-newton' (optimization loop 2), the Schur-decomposition routine, and the sorting of the real Schur-vectors. 

As the state space increases, we observe a quadratic growth in the difference of ETs between cPCCA+ and GenPCCA (see Figures \ref{Fig:ETMacro} and \ref{Fig:ExpFMacro}), while the difference between the coarse-grained transition matrices remains small (see Figure \ref{Fig:DiffMacro}).

\begin{figure}[h!]
	\centering
   	\includegraphics[scale=0.25]{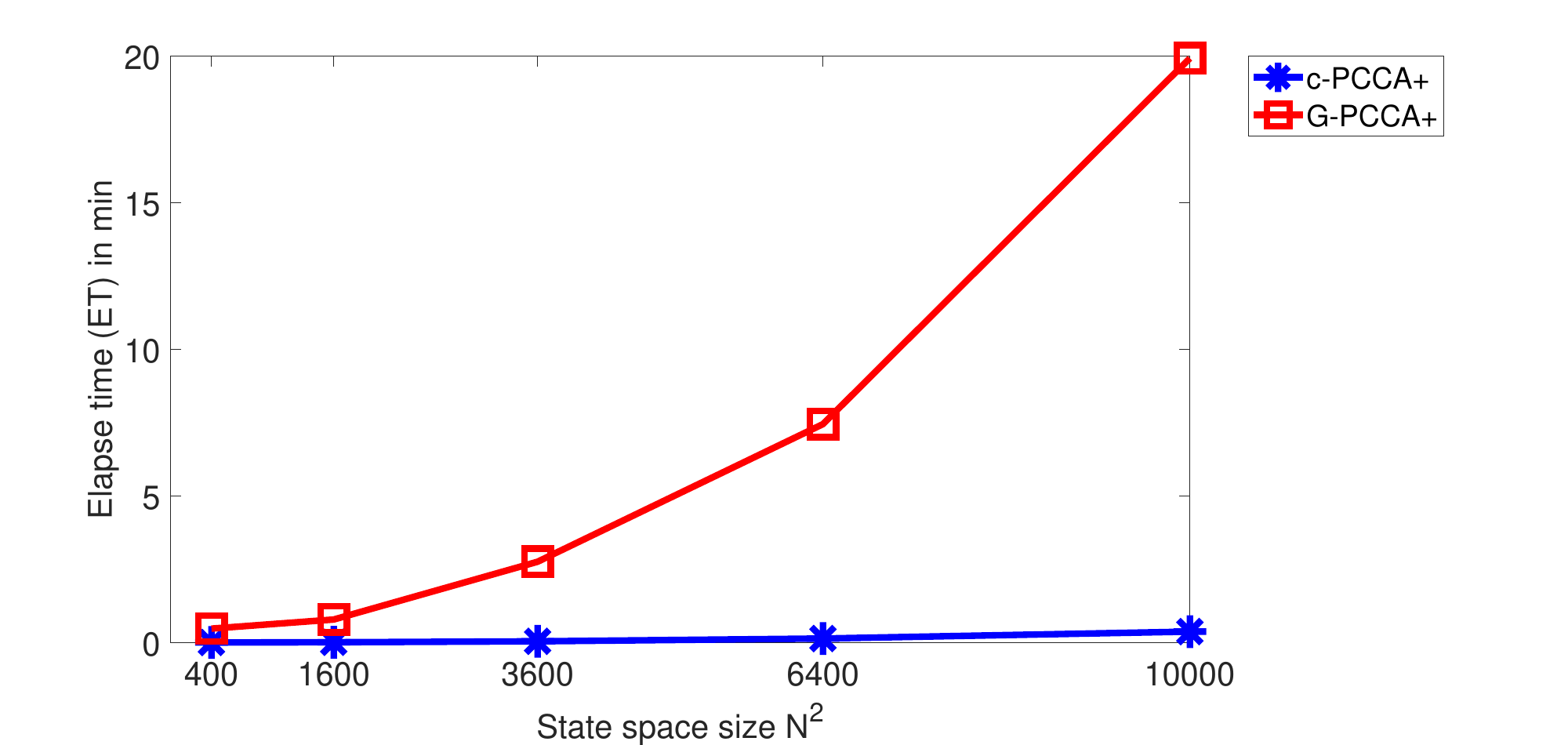}  
\caption{Elapse times (ET) of c-PCCA+ and GenPCCA for the stochastic gene-regulatory network matrix with increasing state space $N^2$. }  
\label{Fig:ETMacro}
\end{figure}

\begin{figure}[h!]
	\centering
		\includegraphics[scale=0.25]{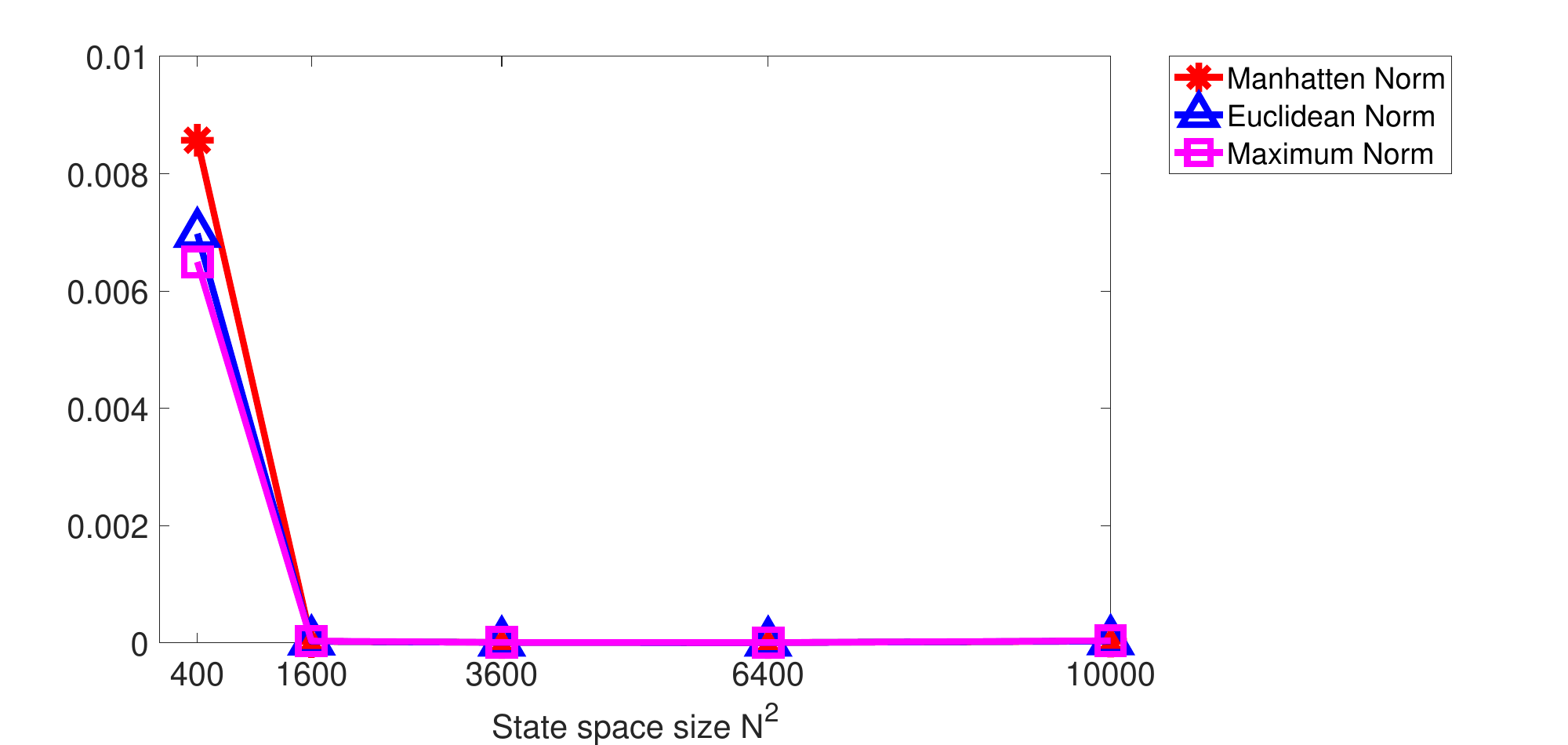}
	  \caption{Normed error differences in  coarse-grained transition matrices $P_c$ calculated with cPCCA+ and GenPCCA for the stochastic gene-regulatory network with increasing state space $N^{2}$.}
	  \label{Fig:DiffMacro}
\end{figure}

We could also observe that the GenPCCA approach spends the majority of the ET performing the Schur-decomposition and sorting the Schur-vectors (see Figure \ref{Fig:SubRMacro}).
\end{enumerate}
With these examples, we have shown that cPCCA+ and GenPCCA are equivalent in that they deliver the same clustering result.

\section{Discussion}\label{Discussion}
This article introduces cPCCA+, an extended and  computationally efficient (fast and accurate) algorithm for coarse-graining of non-reversible stochastic matrices.  
cPCCA+ linearly transforms the complex-valued eigenvectors into real ones, which maintains both the invariant subspace condition and the piecewise constant structure of the eigenvector rows.
In addition to theoretical derivations, the approach was tested on particular examples. 
Numerical experiments show that cPCCA+ has improved runtimes with results equally accurate as those of GenPCCA. Corollary~\ref{Col1} confirms this numerical finding also theoretically. 

The importance of the presented algorithm can be acknowledged by considering the stochastic nature of cellular phenotype switches. 
One prominent example are cancer cells in tumors, which exhibit a diverse phenotype spectrum, and whose transition patterns are assumed to follow a Markov model, see e.g.,  \cite{gupta2011stochastic}.
The original PCCA+ algorithm \cite{roblitz2013fuzzy} has previously been proven valuable for the clustering of Markov chains that describe gene regulatory networks \cite{chu2017markov,tse2018rare}, with its limitation to be only applicable to reversible systems. Our adaptations in cPCCA+ make it now possible to analyze more general systems of, e.g., gene regulatory networks that exhibit  non-reversible transition patterns.  

As demonstrated in this article, both the cPCCA+ and GenPCCA algorithm are suitable for dealing with cases of large non-reversible matrices in Matlab. However, cPCCA+ has been proven to overcome the computational burden that is attributable to  the direct Schur-decomposition and its sorting in the Matlab implementation of GenPCCA.
We point out that there exist non-generic transition matrices $P$ which admit no complex eigen-decompositions but can still be handled by the Schur method \cite{Weber2017}. These cases, however, are not relevant in practice, because very small perturbations in the matrix elements, i.e., smaller than the error resulting from modeling and simulations, turn these matrices into ones that posses an eigen-decomposition.

While we have focused in this article on the PCCA+ algorithms for the Matlab programming language, we would like to point out that the GenPCCA algorithm is also implemented in the Python packages 
\textit{pyGPCCA} \cite{pyGPCCA}
and \textit{cmdtools} \cite{cmdtools}.
The latter introduced the Krylov-Schur-decomoposition \cite{stewart2001krylov} from the PETSc / SLEPc library \cite{STR-7}, which iteratively computes the leading Schur-vectors and therefore overcomes the computational costs of the direct solve and sort routine of the Matlab code.

\section{Conclusion}
The innovation in cPCCA+ lies in the transformation of the complex-valued eigenvectors into a real subspace. By doing so, we extend its application  to  non-reversible stochastic matrices, including circular transition matrices, which often occur in real-world applications such as gene-regulatory networks.
Given its fast runtimes and accurate results, cPCCA+ is an efficient alternative to currently existing approaches in the Matlab programming language.

\newpage
 \appendix
\counterwithin{figure}{section}  

\section{Additional figures for the perturbed circular matrix example (Section 3.1)}

\begin{figure}[h!]
	\centering
		\includegraphics[scale=0.25]{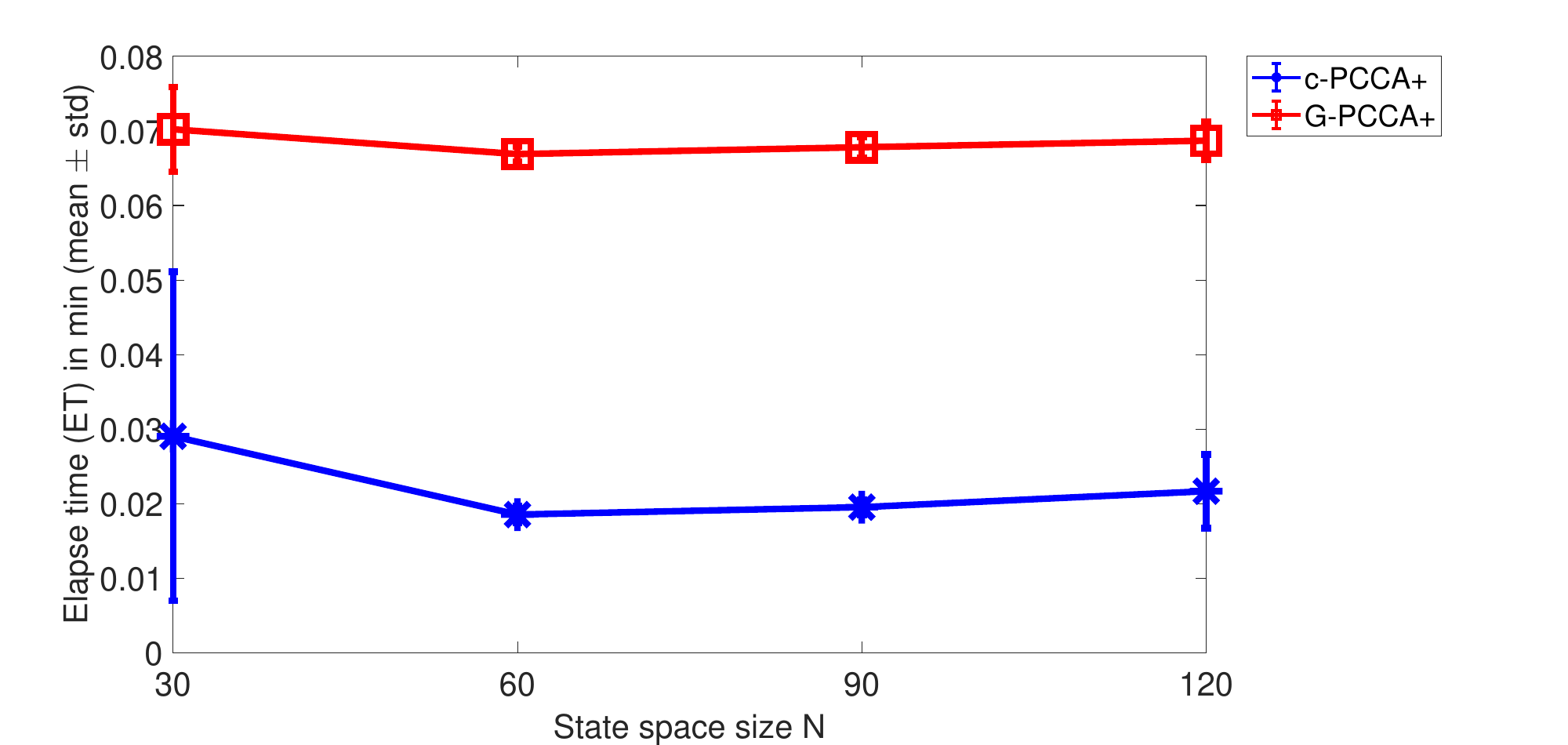}
	  \caption{Elapse Time (ET) of c-PCCA+ and GenPCCA for perturbed circular matrix. The figure shows an error bar plot with the mean and standard deviation of five ETs for each state space size $N$. }
	  \label{Fig:ETCircularPert}
\end{figure}

\begin{figure}[h!]
	\centering
		\includegraphics[scale=0.28]{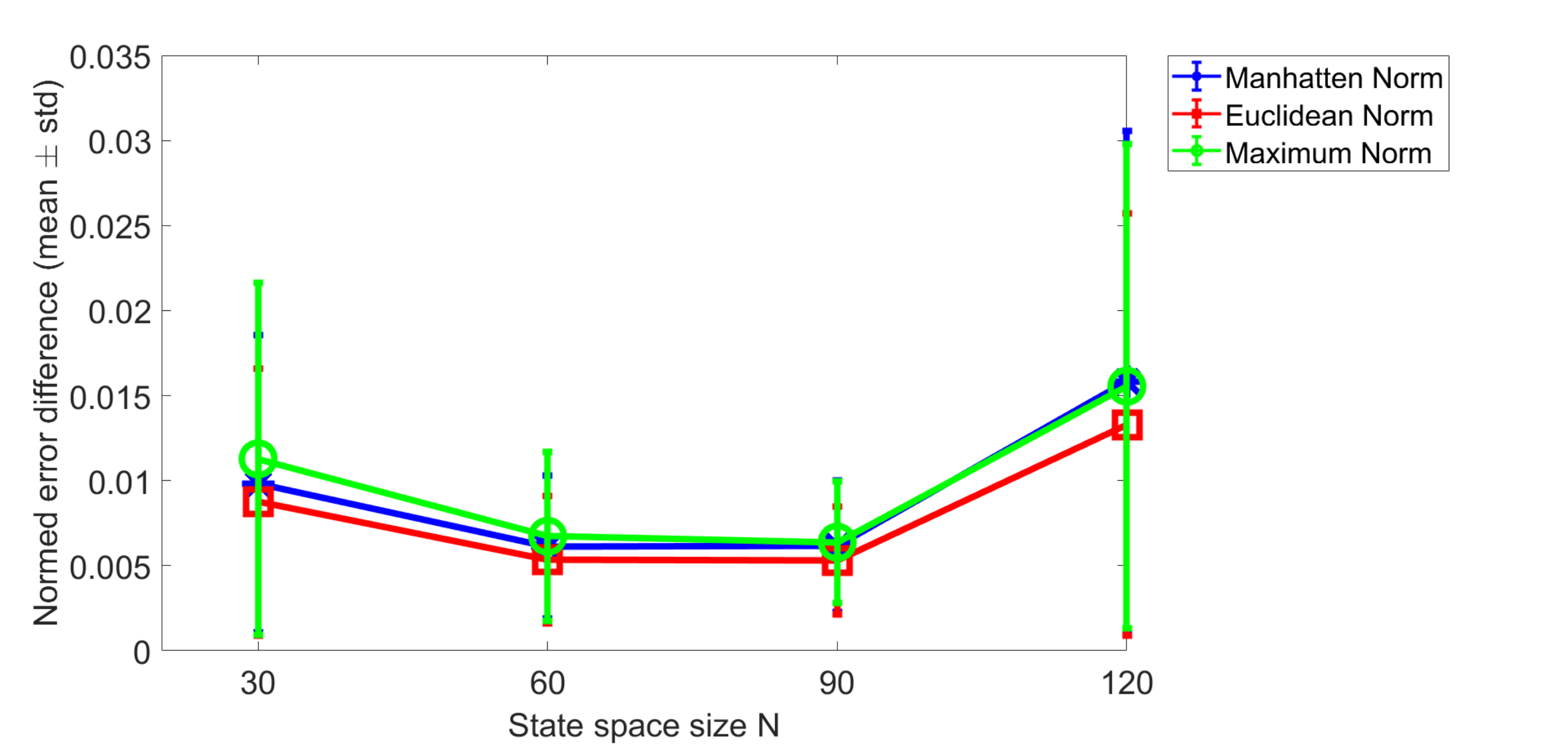}
	  \caption{Normed difference in coarse-grained transition matrices $P_c$ for five perturbed circular matrices per state space size $N$.  }
	  \label{Fig:PertCircDiff}
\end{figure}

\newpage
\section{Additional figures for the gene-regulatory network example (Section 3.2)}
\begin{figure}[h!]
	\centering
   	\includegraphics[scale=0.35]{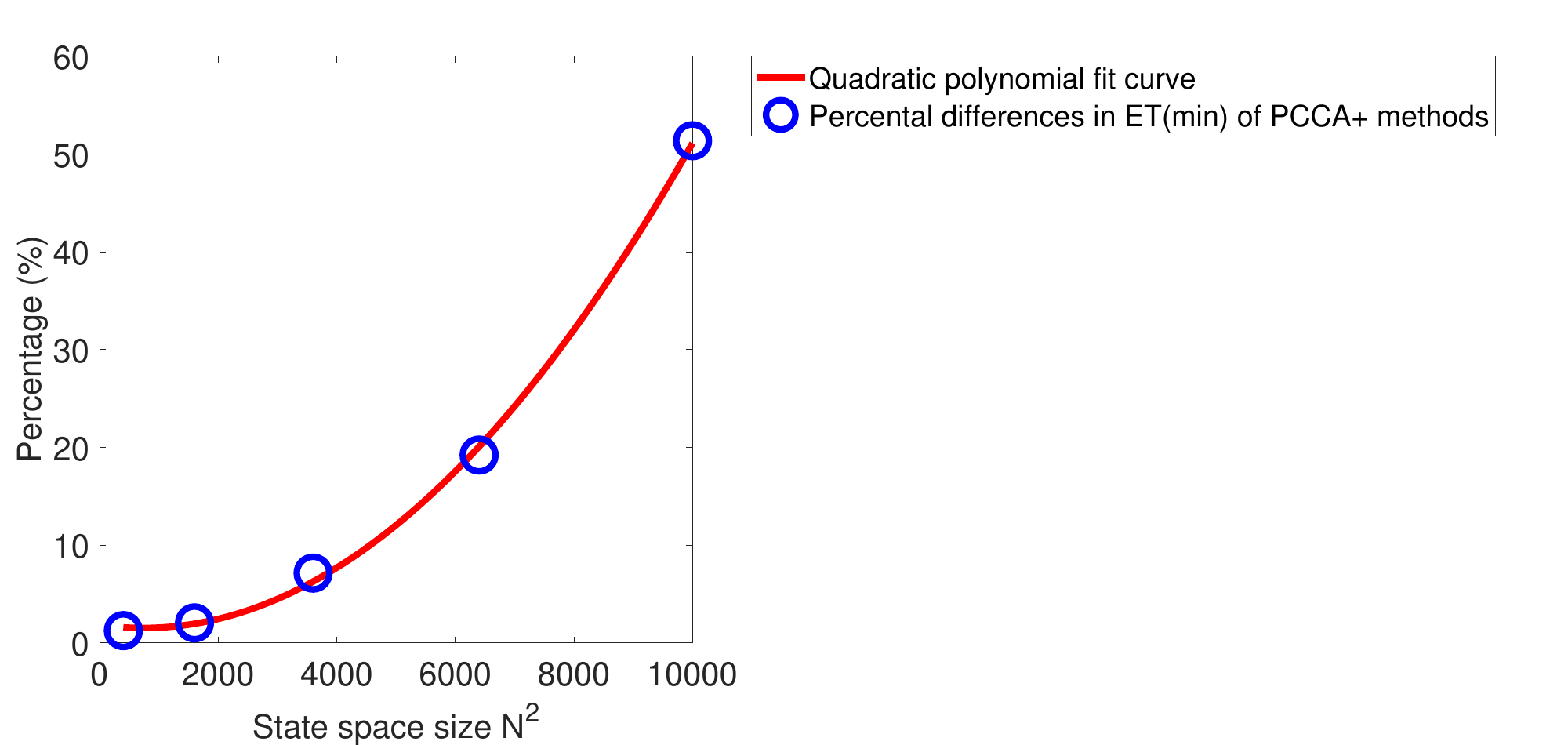} 
\caption{Quadratic polynomial curve fit to percental differences in ET between cPCCA+ and GenPCCA for the gene-regulatory network example for increasing state space $N^2$.}
\label{Fig:ExpFMacro}

\end{figure}

\begin{figure}[h!]
	\centering
		\includegraphics[scale=0.35]{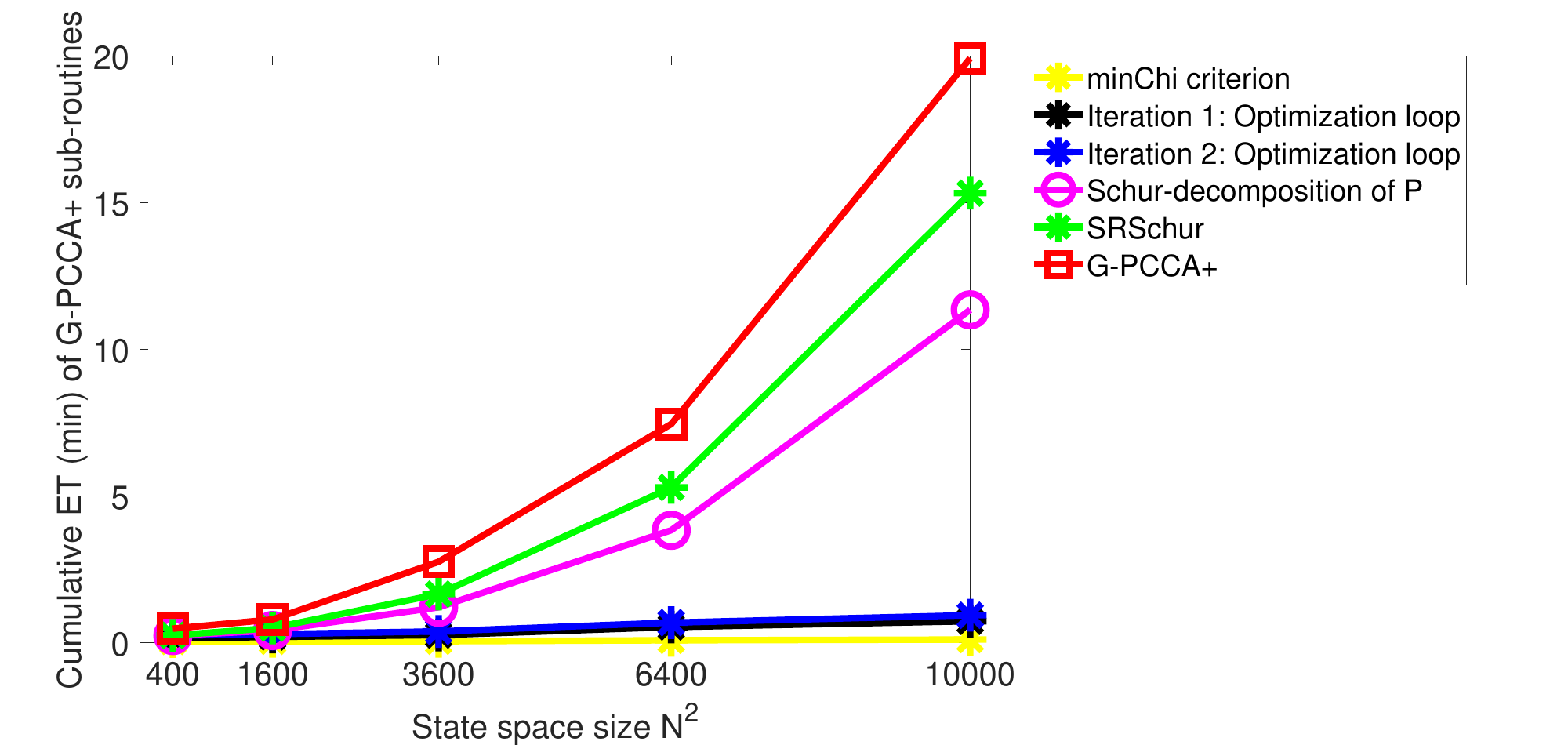}
	  \caption{Cumulative plot of elapse times (ET) for different sub-routines in GenPCCA for increasing state-space $N^2$. Sub-routines are: {\bf minChi criterion:} Determines potential range of numbers of meta-stable states $n_c$; {\bf SRSchur:} Schur-vectors are sorted/re-arranged with respect to their distance form the unit circle; {\bf Optimization loop:} The optimization loop for finding the transformation matrix $A$ is a two-step process. In the first step, the algorithm uses the 'nelder-mead' optimization routine in order to find the optimal cluster number $n_c$ (Iteration 1). In the second step, the 'gauss-newton' routine is used to iteratively optimize $A$ for the found optimal cluster number $n_c$ (Iteration 2). 
	  {\bf Schur decomposition:} Schur decomposition calculates all Schur vectors of matrix $P$. For more information on the different sub-routines, see \cite{reuter2019generalized}
	  }
	  \label{Fig:SubRMacro}
\end{figure}

\newpage
\section*{Data statement}
The cPCCA+ Matlab codes (\textit{cPCCA} \cite{cPCCA2022})  for the examples 1 and 2 can be found on GitHub (\url{https://github.com/sroeblitz/cPCCA/tree/v1.0.0}).
The GenPCCA Matlab code, as described in \cite{reuter2019generalized}, was downloaded from GitHub (\url{https://github.com/msmdev/gpcca}) and adpated to our examples.

\section*{Funding sources}
The work of AF and SR was supported by Trond Mohn Foundation (BSF, https://www.mohnfoundation.no/), grant no. BFS2017TMT01. 
The work of AS has been funded by Deutsche Forschungsgemeinschaft (DFG) through grant CRC 1114 "Scaling Cascades in Complex Systems", Project Number 235221301, Project A05 "Probing scales in equilibrated systems by optimal nonequilibrium forcing".

\section*{Declaration of Competing Interest}
The authors declare that they have no known competing financial interests or personal relationships that could have appeared to influence the work reported in this paper.
\printcredits

\bibliographystyle{unsrt}

\bibliography{cas-refs}

\begin{thebibliography}{10}

\bibitem{chu2017markov}
B.K. Chu, M.J. Tse, R.R. Sato, and E.L. Read.
\newblock {Markov State Models of gene regulatory networks}.
\newblock {\em BMC Syst Biol}, 11(14):1--17, 2017.
\newblock \url{https://doi.org/10.1186/s12918-017-0394-4}.

\bibitem{tse2018rare}
M.J. Tse, B.~K. Chu, C.P. Gallivan, and E.L. Read.
\newblock Rare-event sampling of epigenetic landscapes and phenotype
  transitions.
\newblock {\em PLoS Comput Biol}, 14(8):e1006336, 2018.
\newblock \url{https://doi.org/10.1371/journal.pcbi.1006336}.

\bibitem{swope2004describing}
W.C. Swope, J.W. Pitera, F.~Suits, M.~Pitman, M.~Eleftheriou, B.G. Fitch, R.S.
  Germain, A.~Rayshubski, T.J.C. Ward, Y.~Zhestkov, and R.~Zhou.
\newblock Describing protein folding kinetics by molecular dynamics
  simulations. 2. example applications to alanine dipeptide and a
  $\beta$-hairpin peptide.
\newblock {\em J Phys Chem B}, 108(21):6582--6594, 2004.
\newblock \url{https://doi.org/10.1021/jp037422q}.

\bibitem{noe2009constructing}
F.~No{\'e}, C.~Sch{\"u}tte, E.~Vanden-Eijnden, L.~Reich, and T.R. Weikl.
\newblock Constructing the equilibrium ensemble of folding pathways from short
  off-equilibrium simulations.
\newblock {\em PNAS}, 106(45):19011--19016, 2009.
\newblock \url{https://doi.org/10.1073/pnas.0905466106}.

\bibitem{fersht1995characterizing}
A.R. Fersht.
\newblock Characterizing transition states in protein folding: an essential
  step in the puzzle.
\newblock {\em Curr Opin Struct Biol}, 5(1):79--84, 1995.
\newblock \url{https://doi.org/10.1016/0959-440X(95)80012-P}.

\bibitem{chodera2014markov}
J.D. Chodera and F.~No{\'e}.
\newblock Markov state models of biomolecular conformational dynamics.
\newblock {\em Curr Opin Struct Biol}, 25:135--144, 2014.
\newblock \url{https://doi.org/10.1016/j.sbi.2014.04.002}.

\bibitem{burke2020biochemical}
P.E.P. Burke, C.B.d.L. Campos, L.d.F. Costa, and M.G. Quiles.
\newblock A biochemical network modeling of a whole-cell.
\newblock {\em Sci Rep}, 10(13303):1--14, 2020.
\newblock \url{https://doi.org/10.1038/s41598-020-70145-4}.

\bibitem{reuter2018generalized}
B.~Reuter, M.~Weber, K.~Fackeldey, S.~R{\"o}blitz, and M.E. Garcia.
\newblock {Generalized Markov State Modeling Method for Nonequilibrium
  Biomolecular Dynamics: Exemplified on Amyloid $\beta$ Conformational Dynamics
  Driven by an Oscillating Electric Field}.
\newblock {\em J Chem Theory Comput}, 14(7):3579--3594, 2018.
\newblock \url{https://doi.org/10.1021/acs.jctc.8b00079}.

\bibitem{pande2010everything}
V.S. Pande, K.~Beauchamp, and G.R. Bowman.
\newblock {Everything you wanted to know about Markov State Models but were
  afraid to ask}.
\newblock {\em Methods}, 52(1):99--105, 2010.
\newblock \url{https://doi.org/10.1016/j.ymeth.2010.06.002}.

\bibitem{roblitz2013fuzzy}
S.~R{\"o}blitz and M.~Weber.
\newblock {Fuzzy spectral clustering by PCCA+: application to Markov state
  models and data classification}.
\newblock {\em Adv Data Anal Classif}, 7(2):147--179, 2013.
\newblock \url{https://doi.org/10.1007/s11634-013-0134-6}.

\bibitem{ANDRILLI2010491}
S.~Andrilli and D.~Hecker.
\newblock {Chapter 8 - Additional Applications}.
\newblock In {\em {Elementary Linear Algebra}}, pages 491--585. {Academic
  Press}, {Boston}, {Fourth} edition, 2010.
\newblock \url{https://doi.org/10.1016/B978-0-12-374751-8.00019-6}.

\bibitem{husic2018markov}
B.E. Husic and V.S. Pande.
\newblock {Markov State Models: From an Art to a Science}.
\newblock {\em J Am Chem Soc}, 140(7):2386--2396, 2018.
\newblock \url{https://doi.org/10.1021/jacs.7b12191 }.

\bibitem{fackeldey2018spectral}
K.~Fackeldey, A.~Sikorski, and M.~Weber.
\newblock {Spectral clustering for non-reversible Markov chains}.
\newblock {\em Comp Appl Math}, 37(5):6376--6391, 2018.
\newblock \url{https://doi.org/10.1007/s40314-018-0697-0}.

\bibitem{reuter2019generalized}
B.~Reuter, K.~Fackeldey, and M.~Weber.
\newblock {Generalized Markov modeling of nonreversible molecular kinetics}.
\newblock {\em J Chem Phys}, 150(17):174103, 2019.
\newblock \url{https://doi.org/10.1063/1.5064530}.

\bibitem{brandts2002matlab}
J.H. Brandts.
\newblock {Matlab code for sorting real Schur forms}.
\newblock {\em Numer Linear Algebra Appl}, 9(3):249--261, 2002.
\newblock \url{ https://doi.org/10.1002/nla.274}.

\bibitem{weber2018implications}
M.~Weber.
\newblock {Implications of PCCA+ in Molecular Simulation}.
\newblock {\em Comput}, 6(1):20, 2018.
\newblock \url{ https://doi.org/10.3390/computation6010020}.

\bibitem{deuflhard2005}
P.~Deuflhard and M.~Weber.
\newblock {Robust Perron cluster analysis in conformation dynamics}.
\newblock {\em Linear Algebra Appl}, 398:161--184, 2005.
\newblock \url{https://doi.org/10.1016/j.laa.2004.10.026}.

\bibitem{fackeldey2018spectralZIB}
K.~Fackeldey, A.~Sikorski, and M.~Weber.
\newblock {Spectral Clustering for Non-reversible Markov Chains}.
\newblock Technical Report 18-48, ZIB, 2018.
\newblock Available online at
  \url{https://opus4.kobv.de/opus4-zib/frontdoor/index/index/docId/7021};
  Accessed on June 14, 2022.

\bibitem{frank2021bifurcation}
A.S. Frank, K.~Larripa, H.~Ryu, R.G. Snodgrass, and S.~R{\"o}blitz.
\newblock Bifurcation and sensitivity analysis reveal key drivers of
  multistability in a model of macrophage polarization.
\newblock {\em J Theor Biol}, 509(110511), 2021.
\newblock \url{https://doi.org/10.1016/j.jtbi.2020.110511}.

\bibitem{gupta2011stochastic}
P.B. Gupta, C.M. Fillmore, G.~Jiang, S.D. Shapira, K.~Tao, C.~Kuperwasser, and
  E.S. Lander.
\newblock {Stochastic State Transitions Give Rise to Phenotypic Equilibrium in
  Populations of Cancer Cells}.
\newblock {\em Cell}, 146(4):633--644, 2011.
\newblock \url{https://doi.org/10.1016/j.cell.2011.07.026}.

\bibitem{Weber2017}
M.~Weber.
\newblock {Eigenvalues of non-reversible Markov chains - A case study}.
\newblock Technical Report 17-13, ZIB, 2017.
\newblock Available online at
  \url{https://opus4.kobv.de/opus4-zib/frontdoor/index/index/docId/6219};
  Accessed on June 14, 2022.

\bibitem{pyGPCCA}
B.~Reuter, M.~Klein, and M.~Lange.
\newblock {pyGPCCA -- python GPCCA: Generalized Perron Cluster Cluster Analysis
  package to coarse-grain reversible and non-reversible Markov State Models.}
\newblock GitHub, 2021.
\newblock Available online at \url{https://github.com/msmdev/pyGPCCA}.

\bibitem{cmdtools}
A.~Sikorski, R.~Sechi, and L.~Helfmann.
\newblock cmdtools: Python implementation of severals tools for the analysis of
  dynamical systems from the transfer operator perspective.
\newblock GitHub, 2021.
\newblock \url{https://doi.org/10.5281/zenodo.4749330}; Available online at
  \url{https://github.com/zib-cmd/cmdtools/tree/v1.0.1}.

\bibitem{stewart2001krylov}
G.W. Stewart.
\newblock {A {K}rylov-{S}chur Algorithm for Large Eigenproblems}.
\newblock {\em SIAM J Matrix Anal Appl}, 23(3):601--614, 2001.
\newblock \url{https://doi.org/10.1137/S0895479800371529}.

\bibitem{STR-7}
V.~Hern\'{a}ndez, J.E. Rom\'{a}n, A.~Tom\'{a}s, and V.~Vidal.
\newblock Krylov-{S}chur methods in {SLEP}c.
\newblock Slepc technical report str-7, Scalable Library for Eigenvalue Problem
  Computations, Universitat Polyt\'{e}cnica de Val\'{e}ncia, June 2007.
\newblock \url{https://slepc.upv.es/documentation/reports/str7.pdf}.

\bibitem{cPCCA2022}
A.S. Frank and S.~R{\"o}blitz.
\newblock {cPCCA: Robust Perron Cluster Analysis for Stochastic Matrices with
  Complex Eigenvalues}.
\newblock GitHub, 2022.
\newblock \url{https://doi.org/10.5281/zenodo.6782413}; Available online at
  \url{https://https://github.com/sroeblitz/cPCCA/tree/v1.0.0}.

\end{thebibliography}

\end{document}